%% file: ms.tex
\documentclass[12pt,preprint]{aastex}
\usepackage{amsmath}
\usepackage{bm}
\slugcomment{}

\shorttitle{3D-simulations of SASI in Supernovae}
\shortauthors{Iwakami et al.}

\begin{document}


\title{Three-Dimensional Simulations of Standing Accretion Shock
Instability in Core-Collapse Supernovae }


\author{Wakana Iwakami\altaffilmark{1}, Kei Kotake\altaffilmark{2,3}, Naofumi Ohnishi\altaffilmark{1,4}, Shoichi Yamada\altaffilmark{5,6} 
and Keisuke Sawada\altaffilmark{1}}
 
\affil{$^1$Department of Aerospace Engineering, Tohoku University,
6-6-01 Aramaki-Aza-Aoba, Aoba-ku, Sendai, 980-8579, Japan}
\email{iwakami@rhd.mech.tohoku.ac.jp}
\affil{$^2$Division of Theoretical Astronomy, National Astronomical Observatory of Japan, 2-21-1, Osawa, Mitaka, Tokyo, 181-8588, Japan}
\affil{$^3$ Max-Planck-Institut f\"{u}r Astrophysik, Karl-Schwarzshild
-Str. 1, D-85741, Garching, Germany}
\affil{$^4$Center for Research Strategy and Support, Tohoku University,
6-6-01 Aramaki-Aza-Aoba, Aoba-ku, Sendai, 980-8579, Japan}
%\email{kkotake@th.nao.ac.jp}
\affil{$^5$Science \& Engineering, Waseda University, 3-4-1 Okubo, Shinjuku,
Tokyo, 169-8555, Japan}
\affil{$^6$Advanced Research Institute for Science and Engineering, Waseda University, 3-4-1 Okubo, Shinjuku,
Tokyo, 169-8555, Japan}

\begin{abstract}
We have studied non-axisymmetric standing accretion shock instability, or SASI, by 3D hydrodynamical simulations.
This is an extention of our previous study on axisymmetric SASI. We have prepared a spherically symmetric and steady 
accretion flow through a standing shock wave onto a proto-neutron star, taking into account a realistic equation of state and
neutrino heating and cooling. This unperturbed model is supposed to represent approximately the typical post-bounce 
phase of core-collapse supernovae. We then have added a small perturbation ($\sim$1\%) to the radial velocity and 
computed the ensuing evolutions. Not only axisymmetric but non-axisymmetric perturbations have been also imposed.
We have applied mode analysis to the non-spherical deformation of the shock surface, using the spherical harmonics. 
We have found that (1) the growth rates of SASI are degenerate with respect to the azimuthal index $m$ of the spherical 
harmonics $Y_{l}^{m}$, just as expected for a spherically symmetric background, (2) nonlinear mode couplings produce only $m=0$ modes
for the axisymmetric perturbations, whereas $m \neq 0$ modes are also generated in the non-axisymmetric cases according to  
the selection rule for the quadratic couplings, (3) the nonlinear saturation level of each mode is lower in general for 3D than for 2D 
because a larger number of modes are contributing to turbulence in 3D, (4) low $l$ modes are dominant in the nonlinear phase, 
(5) the equi-partition is nearly established among different $m$ modes in the nonlinear phase, (6) the spectra with respect to 
$l$ obey power laws with a slope slightly steeper for 3D, and (7) although these features are common to the models with 
and without a shock revival at the end of simulation, the dominance of low $l$ modes is more remarkable in the models with a shock revival.
\end{abstract}

\clearpage

\section{Introduction}
Many efforts have been made for the multi-dimensional modeling of
core-collapse supernovae (see \cite{tomas,kotake_rev}
for reviews), urged by accumulated observational evidences that core-collapse supernovae are
globally aspherical commonly \citep{wang96,wang01,wang02}.
Various mechanisms to produce the asymmetry have been discussed so
far: convection (e.g., \citet{herant_94,burrows_95,jankamueller96}), magnetic field
and rapid rotation (see, e.g., \cite{kotake_rev} for collective
references), standing (/stationary/spherical) accretion shock 
instability, or SASI, \citep{blondin_03,scheck_04,blondin_05,ohnishi_1,ohnishi_2,fog06}, 
and g-mode oscillations of protoneutron stars \citep{burr_new}.
Note, however, that most of them have been investigated with two-dimensional
(2D) simulations.

Recently a 3D study on SASI was reported by \citet{blondin_nat}. 
In 2D, the shock deformation by SASI is described with the Legendre polynomials $P_{l}(\theta)$, or 
the spherical harmonics $Y_{l}^{m}(\theta, \phi)$ with $m=0$. Various numerical simulations 
have demonstrated unequivocally that the $l=1$ mode is dominant and a bipolar sloshing
of the standing shock wave occurs with pulsational strong expansions and contractions along the symmetry axis 
\citep{blondin_03,scheck_04,blondin_05,ohnishi_1,ohnishi_2}. 
In 3D, on the other hand, \citet{blondin_nat} observed the dominance of a non-axisymmetric mode
with $l =1$, $m=1$, which produces a single-armed spiral in the later nonlinear phase. They claimed that this 
"spiral SASI" generates a rotational flow in the accretion (see also \cite{blondin_shaw} for 2D computations in the equatorial plane) 
and that it may be an origin of pulsar spins.
 
There are many questions on 3D SASI remaining to be answered yet, however. For example, we are interested in 
how the growth of SASI differs between 3D and 2D among other things. In particular, the change in the saturation properties 
should be made clear. This will be the focus of this paper. Another issue will be the generation of rotation in the accretion flow 
by SASI~\citep{blondin_nat}. Its efficiency and possible correlation with the net linear momentum should be studied more 
in detail and will be the subject of our forthcoming paper~\citep{iwa07}. Incidentally, it is noted that the neutrino heating
and cooling were entirely ignored and the flow was  assumed to be isentropic in \citet{blondin_nat}, but the  
implementation of these physics is helpful for considering the implications not only for the shock revival but also for 
the nucleosynthesis~\citep{kifo}. 

In this paper, we have performed 3D hydrodynamic simulations, employing a 
realistic equation of state \citep{shen98} and taking into account the 
heating and cooling of matter via neutrino emissions and absorptions on
nucleons as done in our 2D studies \citep{ohnishi_1,ohnishi_2}. The inclusion of 
the neutrino heating allows us to discuss how the critical luminosity for 
SASI-triggered explosion could be changed in 3D from those in 2D.
To answer the questions we raised above, we have varied the initial perturbations 
as well as the neutrino luminosity and compared the growth of SASI 
between 2D and 3D in detail, conducting a mode analysis not only for 
the linear phase but also for the nonlinear saturation phase. 

The plan of this paper is as follows: In Section \ref{sec2}, we describe
the models and numerical methods. We show the main numerical results in Section \ref{sec3}.  
We conclude the paper in Section \ref{sec4}. 

\section{Numerical Models\label{sec2}}

The numerical methods employed in the present paper are based on 
the code ZEUS-MP/2 \citep{hayes}, which is a computational fluid dynamics code
for the simulation of astrophysical phenomena, parallelized by the MPI (message-passing) library.
%\footnote{http://lca.ucsd.edu/portal/codes/zeusmp2}. 
The ZEUS-MP/2 code employs the Eulerian hydrodynamics algorithms based on the finite-difference method with a staggered mesh.
In this study, we have modified the original code substantially according to the prescriptions in 
our preceding 2D simulations \citep{ ohnishi_1, ohnishi_2}.

We consider spherical coordinates $(r, \theta, \phi)$ with the origin at the center of the proto-neutron star.
The basic evolution equations describing accretion flows of matter attracted by a proto-neutron star and
irradiated by neutrinos emitted from the proto-neutron star are written as follows,
\begin{equation}
 \frac{d\rho}{dt} + \rho \nabla \cdot \mbox{\boldmath$v$} = 0,
\end{equation}
\begin{equation}
 \rho \frac{d \mbox{\boldmath$v$}}{dt} = - \nabla P - \rho\nabla\Phi - \nabla \cdot \mathbf{Q},
 \label{eq:momentum}
\end{equation}
\begin{equation}
 \rho \frac{d}{dt}\displaystyle{\Bigl(\frac{e}{\rho}\Bigr)}
  = - P \nabla \cdot \mbox{\boldmath$v$} + Q_{\text{E}} - \mathbf{Q} : \nabla \bm{v},
  \label{eq:energy}
\end{equation}
\begin{equation}
 \frac{dY_{\rm e}}{dt} = Q_{\text{N}},
  \label{eq:ye_flow}
\end{equation}
\begin{equation}
 \Phi = - \frac{G M_{\rm in}}{r},
  \label{eq:domain_g}
\end{equation}
where $\rho, \mbox{\boldmath$v$}, e, P, Y_{\rm e}$, and $\Phi$ are the 
density, velocity, internal energy, pressure, electron fraction, and gravitational potential, respectively.
$G$ is the gravitational constant. The self-gravity of matter in the accretion flow is ignored.
$\mathbf{Q}$ is the artificial viscous tensor. $Q_{\rm E}$ and $Q_{\rm N}$ represent 
the heating/cooling and electron source/sink via neutrino absorptions and emissions by free 
nucleons. The Lagrangian derivative is denoted by $d/dt \equiv \partial / \partial t + \mbox{\boldmath$v$} \cdot \nabla$.
The tabulated realistic equation of state based on the relativistic mean field theory \citep{shen98} is
implemented according to the prescription in \citet{kotake}. The mass accretion rate and the mass of the central object are fixed
to be $\dot{M} = 1~M_{\odot}$~s$^{-1}$ and $M_{\rm in} =1.4 M_\odot$, respectively.
The neutrino heating is estimated under the assumptions that neutrinos are emitted isotropically from the central object 
and that the neutrino flux is not affected by local absorptions and emissions (See \citet{ohnishi_1}). 
We consider only the interactions of electron-type neutrinos and anti-neutrinos.
Their temperatures are also assumed to be constant and set to be $T_{\nu_{\text{e}}} = 4 $~MeV and
 $T_{\bar{\nu}_{\text{e}}} = 5 $~MeV, which are the typical values in the post-bounce phase. 
The neutrino luminosity is varied in the range of $L_\nu = 6.0 - 6.8 \times 10^{52}$~ergs~s$^{-1}$.

The spherical polar coordinates are adopted. In the radial direction, the computational mesh is nonuniform,
while the grid points are equally spaced in other directions. We use 300 radial mesh points to cover
$r_{\rm in} \leq r \leq r_{\rm out}$, where $r_{\rm in} \sim 50 ~{\rm km}$ is the radius of the inner boundary 
located roughly at the neutrino sphere and $r_{\rm out} = 2000 ~{\rm km}$ is the radius of the outer boundary, 
at which the flow is supersonic. 30 polar and 60 azimuthal mesh points are used to cover the whole solid angle.
In order to see if this angular resolution is sufficient, we have computed a model with the $300 \times 60 \times 120$ 
mesh points and compared it to the counterpart with the $300 \times 30 \times 60$ mesh points. As shown later in Appendix~\ref{check}, 
the results agree reasonably well with each other both in the linear and nonlinear phases. Although the computational 
cost does not allow us to do the convergence test further, we think that the resolution of this study is good enough.

We use an artificial viscosity of tensor-type described in Appendix~\ref{visc} instead of the von Neumann and Richtmyer-type 
that was originally employed in ZEUS-MP/2. For 3D simulations with the spherical polar mesh,
we find it far better to employ the former than the latter to prevent the so-called carbuncle instability~\citep{quirk}, 
which we observe around the shock front near the symmetry axis, $\theta \sim 0, \ \pi$.  
With the original artifitial viscosity, an appropriate dissipation is not obtained in the azimuthal direction for the shear flow 
resulting from the converging accretion particularly when a fine mesh is used \citep{stone}.
We have applied this artificial viscosity also to axisymmetric 2D simulations and reproduced the previous results~\citep{ohnishi_1}.

\begin{figure}[hbt]
\begin{picture}(0,275)
\put(0,130){\includegraphics[width=55mm]{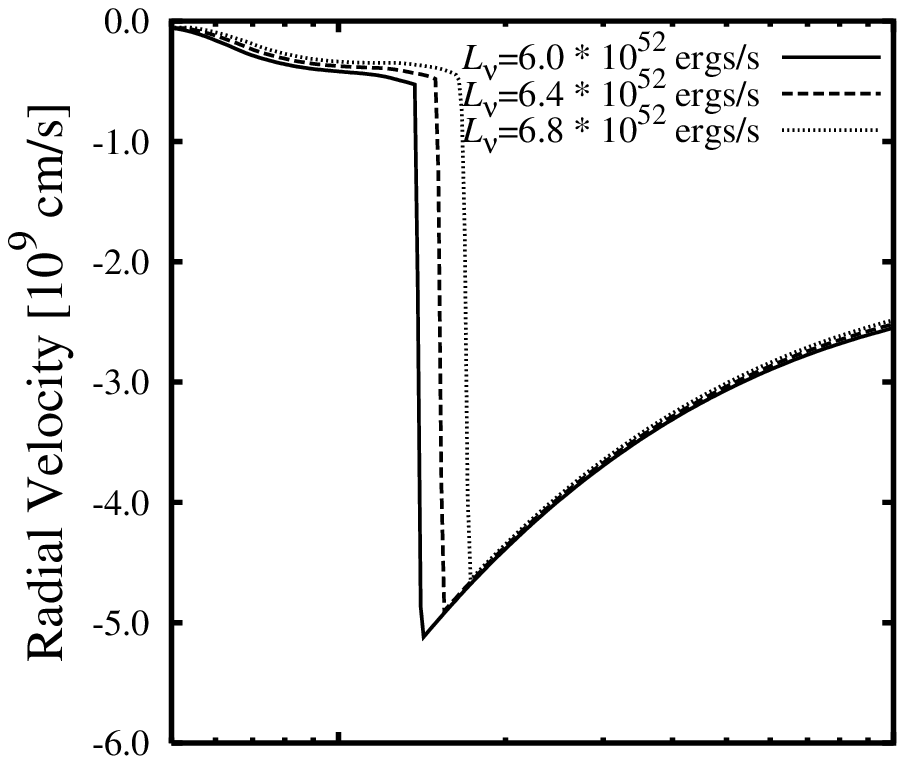}}
\put(155,130){\includegraphics[width=55mm]{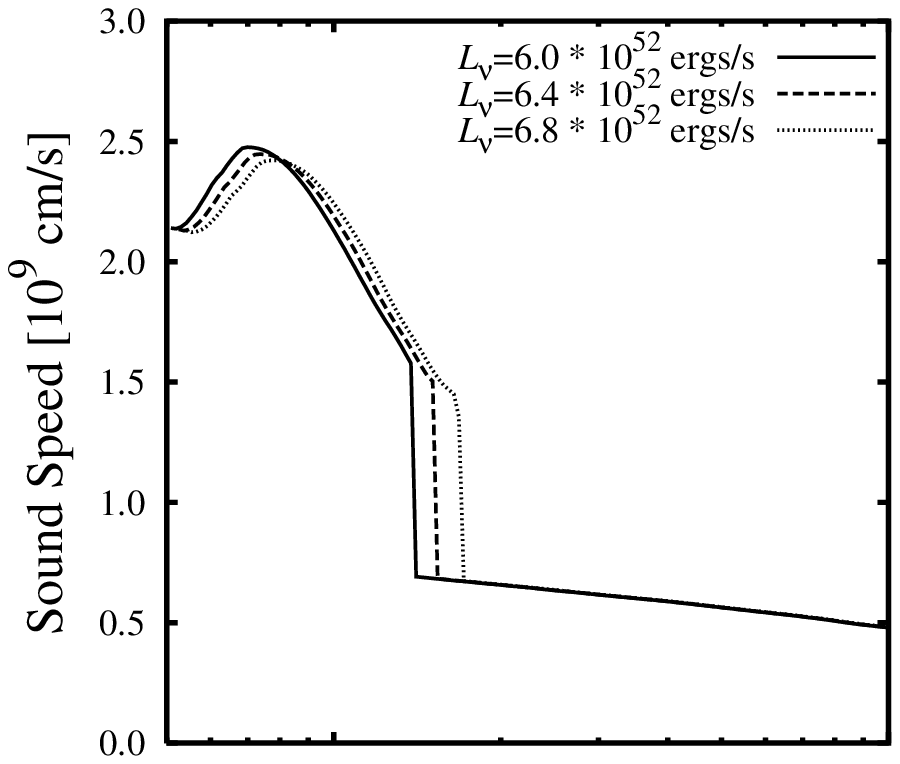}}
\put(310,130){\includegraphics[width=55mm]{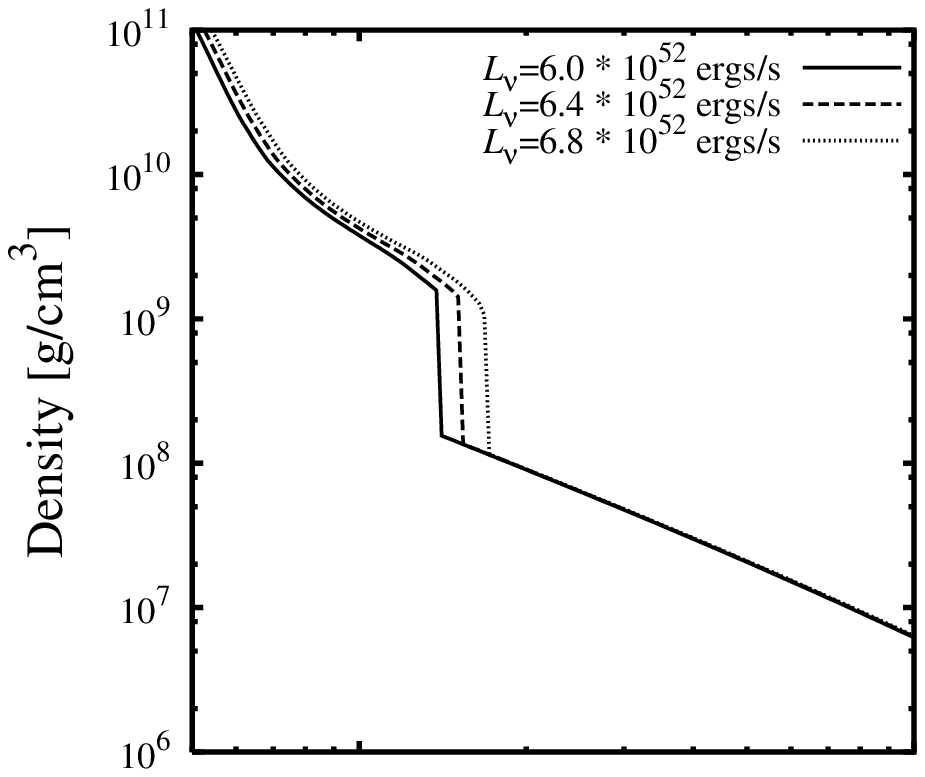}}
\put(0,0){\includegraphics[width=55mm]{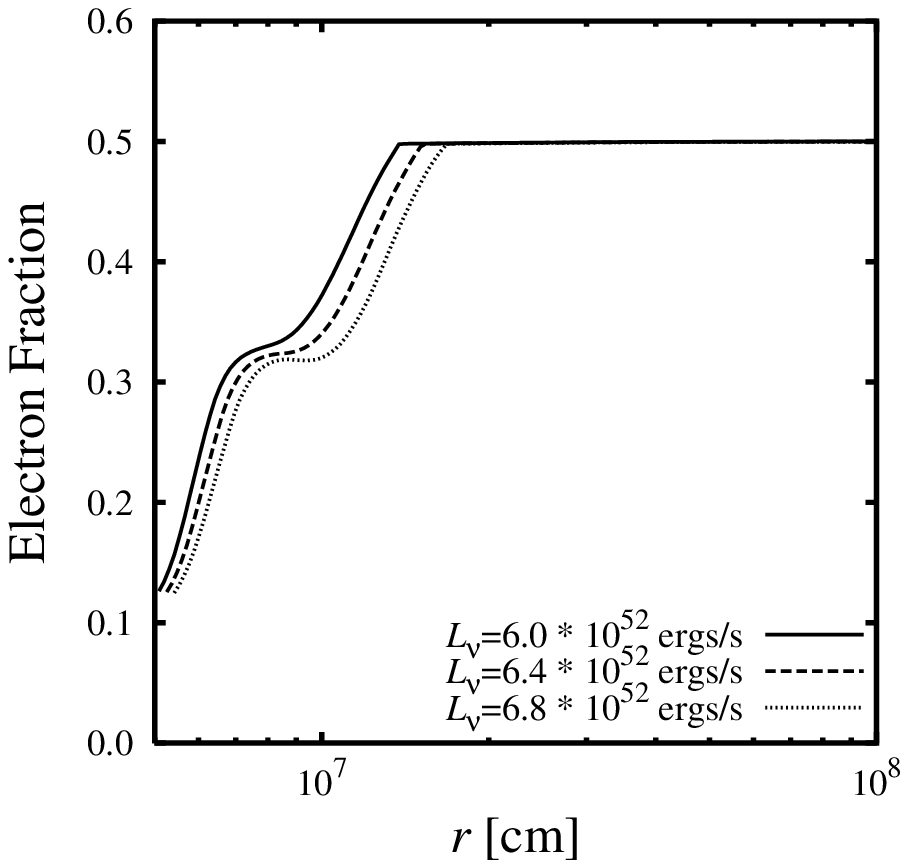}}
\put(155,0){\includegraphics[width=55mm]{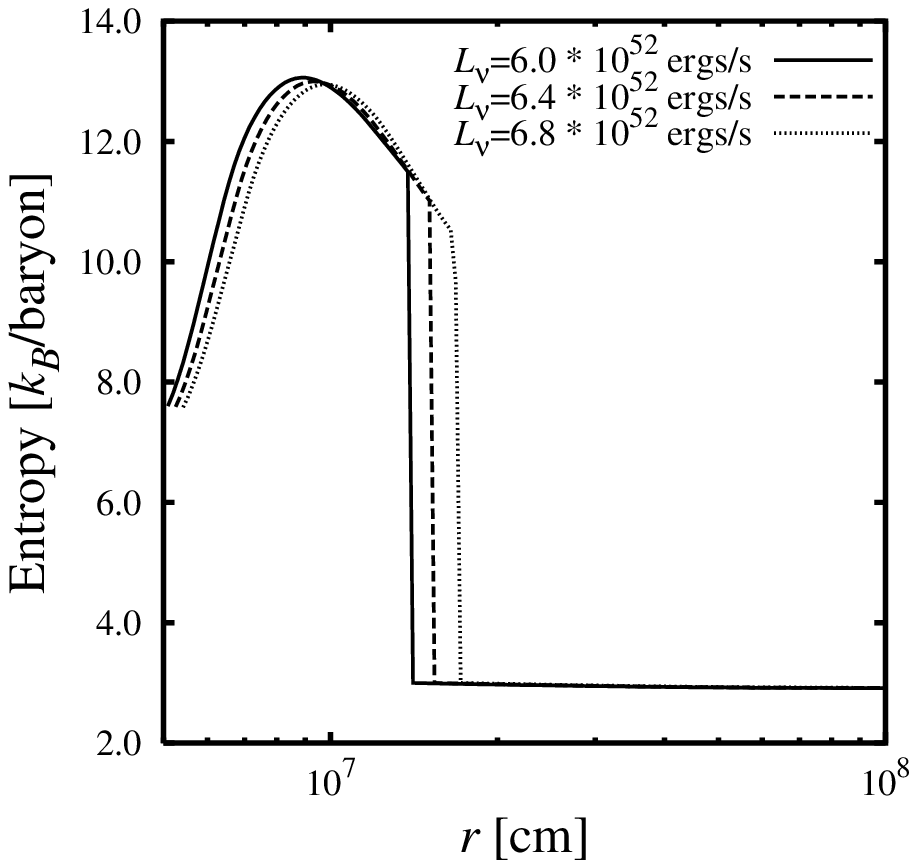}}
\put(310,0){\includegraphics[width=55mm]{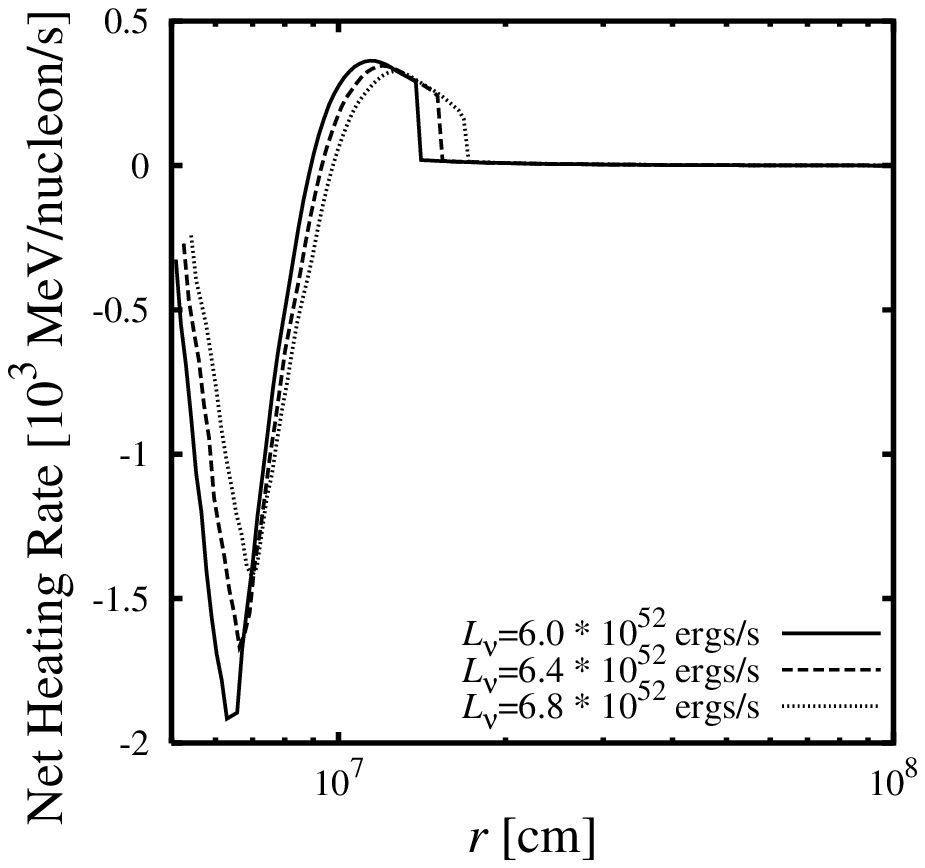}}
\end{picture}
\caption{Profiles of various variables for the unperturbed spherically symmetric accretion flows with 
different neutrino luminosities}
\label{fig1}
\end{figure}

Figure~\ref{fig1} shows the radial distributions of various variables for the unperturbed flows.
The spherically symmetric steady accretion flow through a standing shock wave is prepared in the same manner as in \citet{ohnishi_1}.
Behind the shock wave, the electron fraction is less than 0.5 owing to the electron capture, and a region of negative entropy 
gradient with positive net heating rates is formed for the neutrino luminosities, $L_\nu = 6.0 - 6.8 \times 10^{52}$~ergs~s$^{-1}$.
The values of these variables on the ghost mesh points at the outer boundary are fixed to be constant in time while on the ghost mesh points 
at the inner boundary they are set to be identical to those on the adjacent active mesh points, except for the radial velocity, 
which is fixed to the initial value both on the inner and outer boundaries.

In order to induce the non-spherical instability, we have added a radial velocity perturbation, $\delta v_r(r, \theta, \phi)$, to 
the steady spherically symmetric flow according to the following equation,
\begin{equation}
 v_r(r, \theta, \phi) = v_r^{1D}(r) +\delta v_r(\theta, \phi),
\end{equation}
where $v_r^{1D}(r)$ is the unperturbed radial velocity.
In this study, we consider three types of perturbations:
(1) an axisymmetric ($l=1, m=0$) single-mode perturbation,
\begin{equation}
\delta v_r(\theta, \phi) \propto \sqrt{\frac{3}{4\pi}}\cos\theta \cdot  v_r^{1D}(r),
\end{equation}
(2) a non-axisymmetric perturbation with $l=1$, 
\begin{equation}
\delta v_r(r, \theta, \phi) \propto \left[\sqrt{\frac{3}{4\pi}}\cos\theta  - \sqrt{\frac{3}{8\pi}}\sin\theta\cos\phi \right] \cdot  v_r^{1D}(r),
\end{equation}
and (3) a random multimode perturbation,
\begin{equation}
 \delta v_r(\theta, \phi) \propto {\rm rand} \cdot  v_r^{1D}(r) \quad  (0 \leq {\rm rand} < 1) ,
\end{equation}
where "rand" is a pseudo-random number.
The perturbation amplitude is set to be less than 1\% of the unperturbed velocity. It is noted that there is no
distinction between $m=1$ and $-1$ modes as long as the initial perturbation is added only to the radial velocity, 
as is the case in this paper. To put it another way, the $m=\pm 1$ modes contribute equally. Hence they are referred to as
the $|m|=1$ mode in the following. On the other hand, differences do show up when the initial perturbation is added also to the non-radial 
velocity components, which case is important in discussing the origin of pulsar spins proposed by \citet{blondin_nat} 
and will be the subject of our forthcoming paper~\citep{iwa07}. All the models presented in this paper are summarized in table~\ref{model}.

In the next section, we perform the mode analysis as follows. The deformation of the shock surface 
can be expanded as a linear combination of the spherical harmonics components $Y^m_{l} (\theta, \phi)$:
\begin{equation}
R_S(\theta, \phi) = \sum^{\infty}_{l=0} \sum^{l}_{m=-l} c^m_{l} \, Y^m_{l}(\theta, \phi),
\label{eq:eq10}
\end{equation}
where $Y^m_{l}$ is expressed by the associated Legendre polynomial $P^m_{l}$ and 
a constant $K^m_{l}$ given as
\begin{equation}
Y^m_{l} = K^m_{l} P^m_{l}(\cos \theta) \, e^{im\phi},
\end{equation}
\begin{equation}
K^m_{l} = \sqrt{\frac{2l+1}{4\pi}\frac{(l-m)!}{(l+m)!}}.
\end{equation}
The expansion coefficients can be obtained as follows,
\begin{equation}
c^m_{l}=\int^{2\pi}_0 \! \! \! \! d\phi  \! \int^{\pi}_0 \! \! d\theta \, \sin \theta \, R_S(\theta, \phi) \, Y^{m*}_{l} (\theta, \phi),
\end{equation}
where the superscript * denotes complex conjugation.

\section{Results and Discussions\label{sec3}}
\subsection{Axisymmetric Single-Mode Perturbation\label{sec3_1}}
Before presenting the results of 3D simulations, we first demonstrate the validity of our newly developed 3D code.
We compare the axisymmetric flows obtained by 2D and 3D simulations for the axisymmetric ($l=1, m=0$) 
single-mode perturbation. By 2D simulations we mean that axisymmetry is assumed and computations are done in the meridian section 
with all $\phi$-derivatives being dropped, whereas in 3D simulations we retain all these derivatives and 3D mesh is employed.
This validation is important because numerical errors may induce azimuthal motions even for the axisymmetric initial conditions. Hence we have to confirm that azimuthal errors do not appear in the non-axisymmetric simulation. 

\begin{figure}[hbt]
\begin{picture}(0,200)
\put(10,10){\includegraphics[width=75mm]{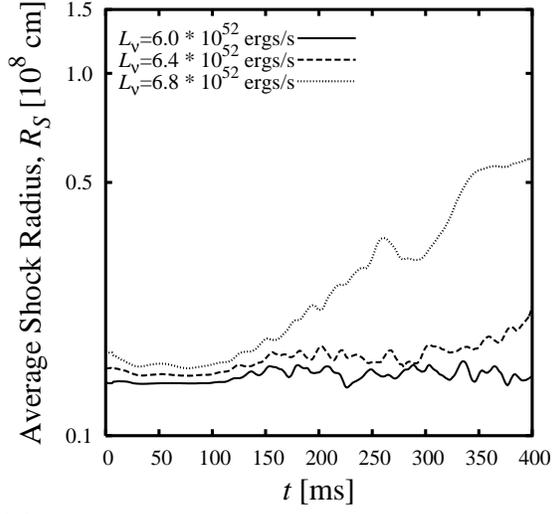}}
\put(10,0){\scalebox{1.1}{\footnotesize (a) 2D axisymmetric perturbation}}
\put(235,10){\includegraphics[width=75mm]{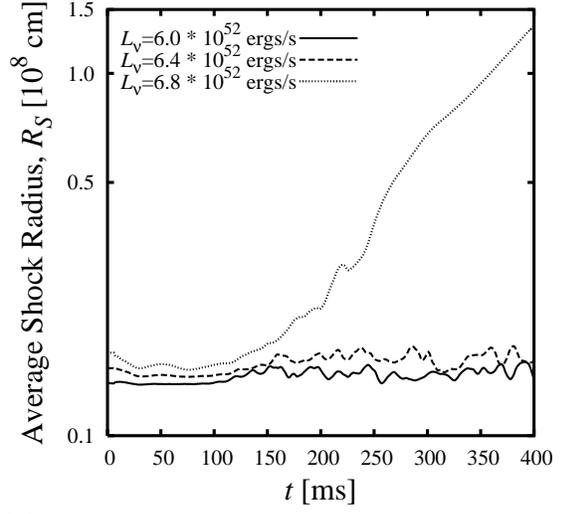}}
\put(235,0){\scalebox{1.1}{\footnotesize (b) 3D axisymmetric perturbation}}
\end{picture}
\caption{The time evolutions of the average shock radius $R_S$ in 2D (left panel) and 3D (right panel) for 
the axisymmetric $l=1, m=0$ single-mode perturbation.}
\label{fig2}
\end{figure}

Figure~\ref{fig2} shows the time evolutions of the average shock radius $R_S$.
The 2D and 3D results are displayed in Figure~\ref{fig2} (a) and (b), respectively.
The average shock radius $R_S$ is obtained from the expansion coefficient $c_0^0$ in Eq.~(\ref{eq:eq10}) multiplied by $K^0_0$.
The solid, broken, and dotted lines correspond to the neutrino luminosities $L_\nu$ of $6.0, 6.4$ and $6.8 \times 10^{52}$ ergs s$^{-1}$, respectively. 
It should be first mentioned that one cannot expect an perfect agreement between two computations of the exponential growth of
the instability followed by the turbulent motions through mode couplings as considered here. It is still obvious that the results of the 2D and 3D
simulations agree on essential features. In particular, both in the 2D and 3D results, we see that the $R_S$ is settled to an almost 
constant value for $L_\nu=6.0$ and $6.4 \times 10^{52}$ ergs s$^{-1}$, whereas it continues to increase for $L_\nu=6.8 \times 10^{52}$ ergs s$^{-1}$.

\begin{figure}[hbt]
\begin{picture}(0,200)
\put(10,10){\includegraphics[width=75mm]{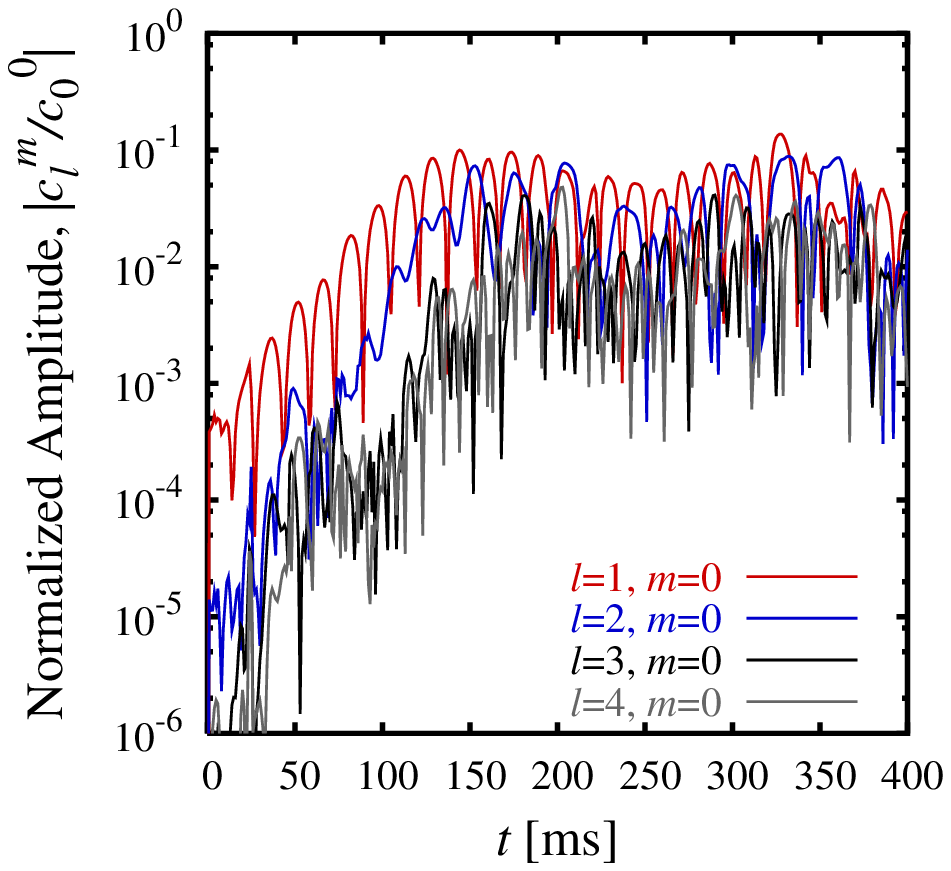}}
\put(10,0){\scalebox{1.1}{\footnotesize (a) 2D axisymmetric perturbation}}
\put(235,10){\includegraphics[width=75mm]{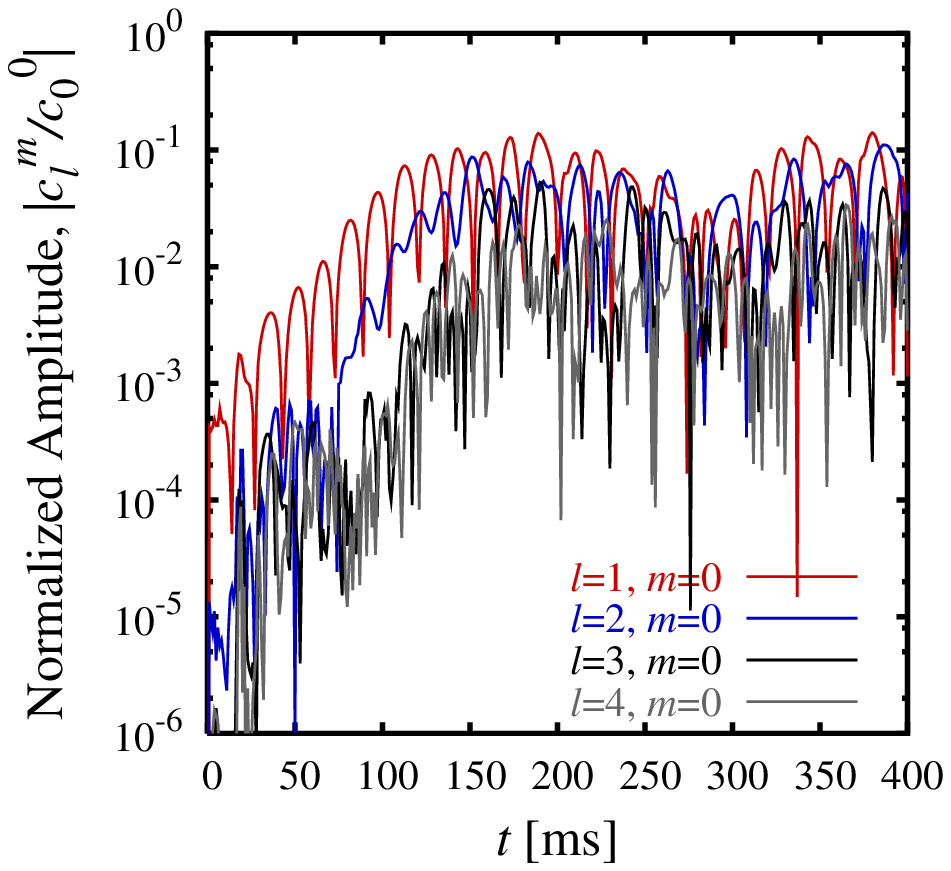}}
\put(235,0){\scalebox{1.1}{\footnotesize (b) 3D axisymmetric perturbation}}
\end{picture}
\caption{The time evolutions of the normalized amplitudes $|c^m_l/c^0_0|$ for Model~I with
the axisymmetric $l=1, m=0$ single-mode perturbation in 2D (left panel) and 3D (right panel).}
\label{fig3}
\end{figure}

Figure~\ref{fig3} presents the normalized amplitudes $|c^m_l/c^0_0|$ as a function of time for Model~I 
($L_\nu = 6.0 \times 10^{52}$ ergs s$^{-1}$). The 2D and 3D results are displayed in Figure~\ref{fig3} (a) and (b), respectively.
The red, blue, black, and gray solid lines correspond to the modes of $(l, m) =(1, 0), (2, 0), (3, 0)$ and $(4, 0)$, respectively.
As can be seen, the time evolution can be divided into two phases. One is the linear growth phase, in which the amplitude of the 
initially imposed mode with $(l, m) = (1, 0)$ grows exponentially. It lasts for $\sim$ 100 ms. It is also found that higher modes are generated by mode couplings 
and grow exponentially during this phase. Then starts the second phase, which is characterized by the saturation of amplitudes of the order of 0.1.
In this phase, the accretion flow becomes turbulent. It is interesting to note that in this nonlinear saturation phase the amplitude of the 
mode with $(l, m) = (2, 0)$ is almost of the same order as the initially imposed mode with $(l, m) = (1, 0)$ and is dominant over other 
modes, which fact was observed in \citet{ohnishi_1}. 

Most important of all, however, is the fact that all the $m \neq 0$ modes are not produced, implying that the perturbed flow retains axisymmetry, which is a necessary condition for the numerical study of non-axisymmetric instability. 
Although the results of the 2D and 3D simulations are not identical again, the essential features such as the linear growth rate of
the initial perturbation ($l=1, m=0$) and the production of other modes via nonlinear mode couplings as well as the saturation levels in the
nonlinear phase are all in good agreement for the two cases. The quantitative differences between the 2D and 3D results mainly  
originate from the difference in the time steps, which depend on the minimum grid width. Note that in addition to the 300 radial and 30 polar mesh points used 
for the 2D computations, 60 azimuthal mesh points are employed in the 3D simulatioins and, as a result, smaller time steps are usually taken for 
the 3D computations. 

\subsection{Non-Axisymmetric Perturbation with $l = 1$ \label{sec3_2}}
Now we discuss the results of genuinely 3D simulations, in which the non-axisymmetric perturbation with ($l=1, |m| =1$) is added to the axisymmetric 
($l=1, m=0$) perturbation.

\begin{figure}[hbt]
\begin{picture}(0,200)
\put(110,0){\includegraphics[width=75mm]{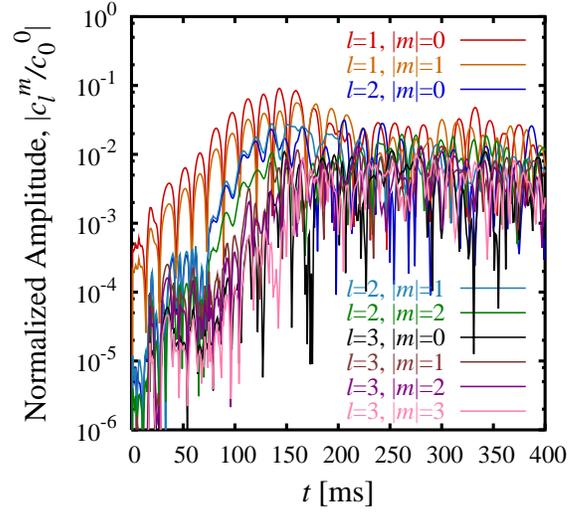}}
\end{picture}
\caption{The time variations of the normalized amplitudes $|c^m_l/c^0_0|$ for Model I\kern-.1emI, in which 
the modes with $(l, |m|) = (1, 0), (1, 1)$ are initially imposed.}
\label{fig4}
\end{figure}

Figure~\ref{fig4} shows the time evolutions of some of the normalized amplitudes $|c^m_l/c^0_0|$ for Model~I\kern-.1emI with the neutrino luminosity 
$L_\nu=6.0 \times 10^{52}$ ergs s$^{-1}$. The red, yellow, blue, light blue, green, black, brown, violet, and pink solid lines denote the amplitudes of the modes with $(l, |m|) = (1,0), (1, 1), (2, 0), (2, 1), (2, 2), (3, 0), (3, 1), (3, 2)$ and $(3, 3)$, 
respectively. 
In the linear phase, the initially imposed modes with $(l, |m|) = (1, 0)$ and $(1, 1)$ grow exponentially just as in the axisymmetric single-mode perturbation. It should be noted that the
growth rate of $(l, |m|) = (1, 1)$ mode is identical to that of $(l, m) = (1, 0)$ mode. This is just as expected
for the spherically symmetric background, for which the modes with the same $l$ but different $m$ are degenerate in the linear eigen frequency. 

It is observed in the figure that the modes with $(l, |m|) =(2, 0), (2, 1), (2, 2)$ are first produced by the nonlinear mode-couplings of   
$(l, |m|) = (1, 0), (1, 1)$. Then the modes with $(l, |m|) = (3, 0), (3, 1), (3, 2), (3, 3)$ can be produced by the couplings of the first-order 
modes with $(l, |m|) = (1, 0), (1, 1)$ and the second-order modes with $(l, |m|) = (2, 0), (2, 1), (2, 2)$.
Even higher order modes are produced subsequently toward the non-linear saturation but are not shown here. 
Although the branching ratios should be investigated more in detail before 
giving any conclusion, the above sequence of the mode generations 
strongly suggests that the nonlinear coupling is mainly of quadratic nature. 

\begin{figure}[hbt]
\begin{picture}(0,200)
\put(30,0){\includegraphics[width=140mm]{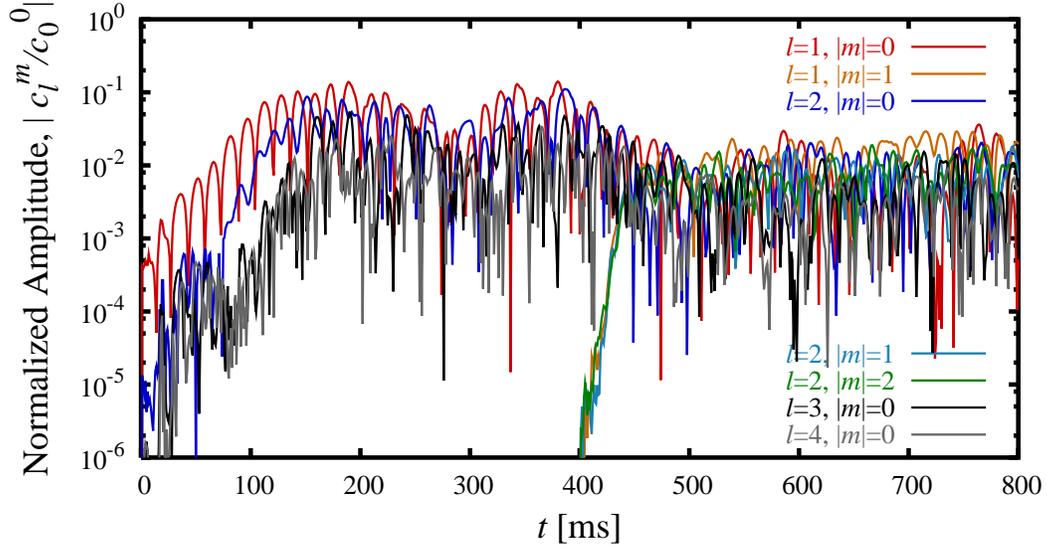}}
\end{picture}
\caption{The time evolutions of the normalized amplitudes $|c^m_l/c^0_0|$ for Model~I\kern-.1emI\kern-.1emI, in which
we have added the random perturbation to the radial velocity at $t=400$ms.}
\label{fig5}
\end{figure}

In the nonlinear phase that begins at $t \sim 200$ms, these mode amplitudes are saturated and settled into a quasi-steady state. As in the axisymmetric case,
the $l = 1$ modes both with $m=0$ and $|m|=1$ are dominant over other modes in this stage for the non-axisymmetric case. 
One thing to be noted here, however, is the fact that the saturated amplitudes for Model~I\kern-.1emI are lower than those for Model~I in general 
(see Figure~\ref{fig3} (b)). This is, we think, because the turbulent energy is shared by a larger number of modes in the non-axisymmetric case 
than in the axistmmetric case. To demonstrate this more clearly, we have added the random perturbation to the radial velocity in the axisymmetric 
Model I at $t=400$ ms, by which time the axisymmetric nonlinear turbulence has fully grown. We refer to this model as Model~I\kern-.1emI\kern-.1emI. 
The entire time evolutions of the normalized amplitudes $|c^m_l/c^0_0|$ for Model~I\kern-.1emI\kern-.1emI are shown in Figure~\ref{fig5}.
It is clear that the axisymmetric $m = 0$ mode amplitudes are reduced after the additional perturbation is given and the non-axisymmetric $m \neq 0$ modes 
grow to the saturation level at $t \sim 450$ ms. The power spectra of the turbulent motions will be discussed more in detail later.

%------------------------------------------------
\subsection{Random Multi-Mode Perturbations \label{random}}

\subsubsection{Dynamical Features and Critical Luminosity\label{basic}}

Having understood the elementary processes of the linear growth and nonlinear mode couplings in the previous section, we  will now discuss the 3D SASI induced 
by the random multimode perturbations, which are supposed to be closer to the reality. In this subsection, we will show the typical dynamics, paying attention to the 
time evolution of the shock wave and, hence, to the critical luminosity, at which the stalled shock is revived.

\begin{figure}[htb]
\begin{picture}(0,200)
\put(10,10){\includegraphics[width=75mm]{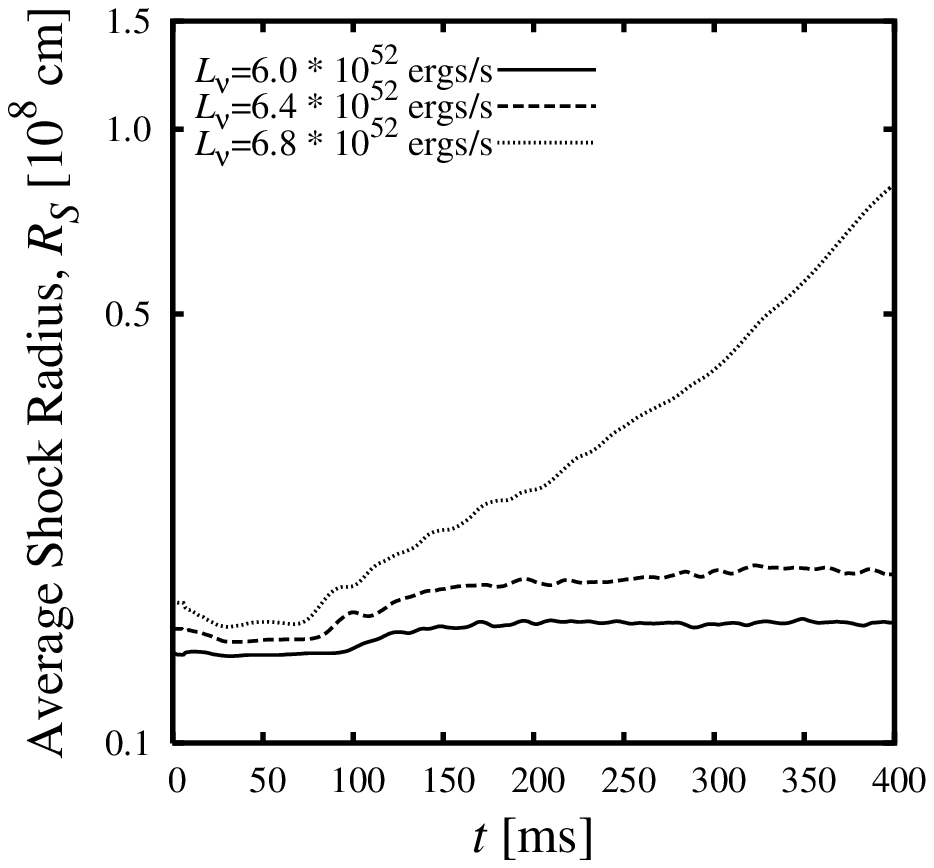}}
\put(10,0){\scalebox{1.1}{\footnotesize (a) 3D non-axisymmetric random perturbation}}
\put(235,10){\includegraphics[width=75mm]{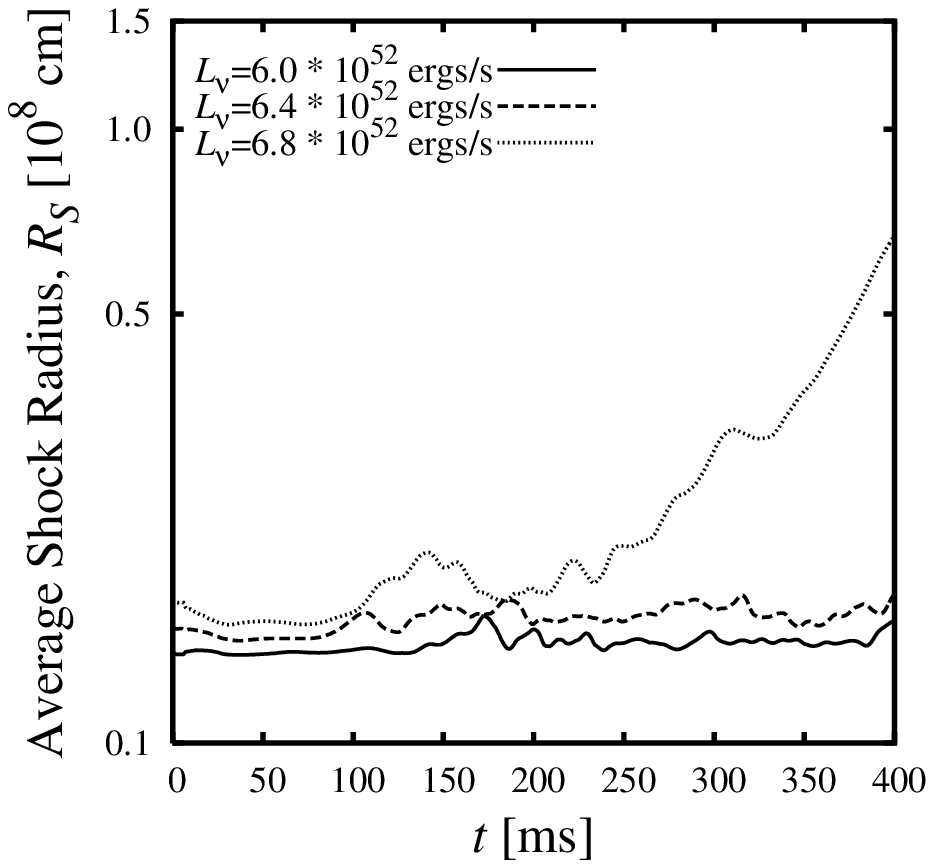}}
\put(235,0){\scalebox{1.1}{\footnotesize (b) 2D axisymmetric random perturbation}}
\end{picture}
\caption{The time evolutions of the average shock radius $R_S$ for Models I\kern-.1emV, V and V\kern-.1emI, in which the random
multi-mode perturbation is imposed in 3D (left panel) and 2D (right panel).}
\label{fig6}
\end{figure}

Figure~\ref{fig6} shows the time variations of the average shock radius $R_S$ for Models I\kern-.1emV, V and V\kern-.1emI with the neutrino luminosities, 
$L_\nu = 6.0, 6.4$ and $6.8 \times 10^{52}$ ergs s$^{-1}$, respectively. It is found that for $L_\nu = 6.0$ and $6.4 \times 10^{52}$ ergs s$^{-1}$, 
the shock is settled to a quasi-steady state after the linear growth whereas it continues to expand for $L_\nu = 6.8 \times 10^{52}$ ergs s$^{-1}$, which is also true for 
the axisymmetric counterpart displayed in the right panel of the figure for comparison. This implies that the critical neutrino luminosity is not much affected by
the change from 2D to 3D. In the following we refer to the models with $L_\nu = 6.0$ and $6.4 \times 10^{52}$ ergs s$^{-1}$ as the non-explosion models and 
the model with $L_\nu =6.8 \times 10^{52}$ ergs s$^{-1}$ as the explosion model and look into their dynamical features in turn. The mode analyses will be done in 
sections \ref{mode_non} and \ref{mode_ex}.

Figure~\ref{fig7} shows the sideviews of the iso-entropy surfaces together with the velocity vectors in the meridian section at four different times for the 
non-explosion model with $L_\nu = 6.0 \times 10^{52}$ ergs s$^{-1}$ (Model I\kern-.1emV). The hemispheres ($0\le \phi \le \pi$) of eight iso-entropy surfaces  
are superimposed on one another and the outermost surface almost corresponds to the shock front. The initial perturbations grow exponentially in the linear phase ((a) $t=40$ ms).
At the end of the linear phase, a blob of high entropy is formed and grows subsequently ((b) $t=93$ ms). High entropy blobs are continuously generated and the nonlinear phase 
begins ((c) $t=100$ ms). Circulating flows occur inside the blobs while high-velocity accretions onto a proto-neutron star surround the blobs. 
These blobs expand and shrink repeatedly, being distorted, splitted and merged with each other inside the shock wave ((d) $t=350$ ms).
Reflecting these complex motions, the shock surface oscillates in all directions but remains alomost spherical for the non-axisymmetric model. This is in sharp contrast to 
the axisymmetric case, in which large-amplitude oscillations occur mainly in the direction of the symmetry axis (\citet{blondin_03, ohnishi_1}).

Figure~\ref{fig8} displays the snapshots of the iso-entropy surfaces and the velocity vectors for the explosion model with $L_\nu = 6.8 \times 10^{52}$ ergs s$^{-1}$ 
(Model V\kern-.1emI). As in the non-explosion model, the sequence of events starts with the linear growth of the initial perturbation inside the shock wave ((a) $t=40$ ms).
In the explosion model, however, many high entropy blobs emerge simultaneously much earlier on ((b) $t=70$ ms) than in the non-explosion model.
Then, these blobs repeat expansions and contractions, being distorted, splitted and merged together just as in the non-explosion counterpart ((c) $t=80$ ms).
After that, some of the blobs get bigger, absorbing other blobs ((d) $t=350$ ms) and, as a result, the shock radius increases almost continuously up to the end of the computation ($t=400$ ms) 
as already demonstrated in Figure~\ref{fig6} (a). At this point of time, the shock surface looks ellipsoidal rather than spherical. However, the major axis is not necessarily aligned with
the symmetry axis and the flow is not symmetric with respect to this major axis, either.

\subsubsection{Mode Spectra \label{mode_non}}

Next we look into the spectral intensity more in detail. As a standard case, we take Model I\kern-.1emV with $L_\nu=6.0 \times 10^{52}$ ergs s$^{-1}$.
Figure~\ref{fig9} shows the time evolutions of the normalized amplitudes $|c_l^m/c^0_0|$ with $m=0$ and compares them with the axisymmetric 
counterparts in Model V\kern-.1emI\kern-.1emI. Note that the $m \neq 0$ modes also exist in the non-axisymmetric model, which are not shown 
in the figure but will be discussed in the following paragraphs. As can be clearly seen, the amplitude of each mode grows exponentially 
until $\sim$ 100 ms, which is the linear phase. Note in particular that the growth rate and oscillation frequency for the $l=1$ mode are the same as those
obtained for the model with the single mode perturbation in the previous section. After $\sim$ 100 ms, the mode amplitudes are saturated and the evolution enters the 
nonlinear phase with a clear dominance of the modes with $l=1, 2$. It is also evident in the figure that the saturation level is lower in general for the non-axisymmetric 
case than for the axisymmetric one, which confirms the claim in the previous section based on the results for the single-mode perturbation.

In Figure~\ref{fig10}, we display the snapshots at four different times of the spectral distributions both for the non-axisymmetric (the left panels) and 
axisymmetric (the right panels) cases. The uppermost panels correspond to the linear phase and the intensity is distributed rather uniformly over all modes.
As the time passes, however, the amplitudes of low $l$ modes grow much more rapidly than those with higher $l$'s (the second and third rows in the figure). 
One can see a similarity in the time evolutions between the non-axisymmetric and axisymmetric models.
After the nonlinear phase starts (the lowermost panels), the growths of all modes are saturated and the spectral distributions are settled to be quasi-steady. 
It is again obvious in these panels that the low $l$ modes are dominant in the nonlinear phase and the saturation level is lower in the non-axisymmetric case.

Now we turn our attention to the quasi-steady turbulence in the nonlinear phase. Shown as a function of $l$ in Figure~\ref{fig11} are the power spectra 
$|c_l^m/c_0^0|^2$ averaged over the interval from $t=150$~ms to 400 ms. Both the non-axisymmetric and axisymmetric cases are displayed by the symbols, 
$\circ$ and $\bullet$, respectively. In the left panel, modes with different $m$'s are plotted separately whereas they are summed up in the right panel.
It is found that the time-averaged power spectra are not much different among the modes with the same $l$ but with a different $m$.
This implies that the equi-partition of the turbulence energy is nearly established among the modes with the same $l$. This will have an important 
ramification for the origin of pulsar spin and will be discussed in our forthcoming paper~\citep{iwa07}. 

The right panel of the figure demonstrates that the time-averaged power spectrum summed over $m$ for the non-axisymmetric case is not much different from 
that for the axisymmetric counterpart. This means that the total turbulence energy does not differ very much between the axisymmetric and non-axisymmetric cases.
As a result, the power of each mode in the non-axisymmetric case is smaller roughly by a factor of $2l + 1$ than that of the same $l$ mode in the axsymmetric case. 
This is the reason why we have observed that the saturation level of the nonlinear SASI is smaller in the non-axisymmetric case than in the axisymmetric case and the shock surface
oscillates in all directions with smaller amplitudes in the non-axisymmetric flow whereas it sloshes in the direction of the symmetry axis with larger amplitudes in the axisymmetric case.
The difference in the fluctuations of the average shock radius seen in Figure~\ref{fig6} (a) and (b) is also explained in the same manner. 

Another interesting feature observed in the figure is the fact that the time-averaged power spectra obey a power-law at $l \gtrsim 10$ for both 
the non-axisymmetric and axisymmetric models. Two straight lines in Figure~\ref{fig11} (a) are the fits to the data for $l \ge 10$, each obtained for the axisimmetric and 
non-axisymmetric models. The powers are found to be $-5.7$ and $-4.3$ for the non-axisymmetric and axisymmetric cases, respectively. Although the difference is 
almost unity, which originates from the difference in multiplicity of the modes with the same $l$, the slope excluding this effect is still a little bit steeper in the non-axisymmetric case
as seen in Figure~\ref{fig11} (b). At the moment we do not know if this difference in slope is real or not and the power itself is also remaining to be explained theoretically.

\subsubsection{The Explosion Models\label{mode_ex}}

So far we have been discussing the non-explosion models, in which the SASI is saturated in the non-linear phase and is settled to a quasi-steady state.  In considering the 
possible consequences of SASI after the supernova explosion occurs, however, it is also interesting to investigate the explosion models, in which a shock revival occurs as a result of SASI. 
 
Figure~\ref{fig12} shows the time evolutions of the normalized amplitudes $|c_l^m/c^0_0|$ for the explosion model (Model V\kern-.1emI) and compares it with the
axisymmetric counterpart (Model V\kern-.1emI\kern-.1emI\kern-.1emI). 
The feature of the explosion models common to the axisymmetric and non-axisymmetric cases is as follows: 
The nonlinear stage is devided into two phases. The earlier phase is quite similar to the nonlinear phase that we have seen so far for the non-explosion models.
The later phase, on the other hand, has a distinctive feature that the $l=1$ mode becomes much more remarkable than other modes and the oscillation period tends to be longer 
as the shock radius gets larger and the advection-acoustic cycle, supposedly an underlying mechanism of the instability, takes longer. The later prominence of $l=1$ mode is 
intriguing and is important to discuss the kick velocity of pulsar quantitatively. The theoretical account, however, is yet to be given.  
The differences between the non-axisymmetric and axisymmetric models, such as the saturation level, are just taken over to the explosion models. 

The power spectra averaged over the time intervals of $100 \le t \le 200$~ms and $250 \le t \le 400$~ms are presented in Figure~\ref{fig13} (a) and (b), respectively.
As in the non-explosion models, the equi-partition among different $m$ modes is again established for the explosion models. This is true even in the later nonlinear phase as seen in 
the right panel of the figure. As a result, the saturation level is smaller for the non-axisymmetric case than for the axisymmetric case as mentioned just above. The power spectra in 
the earlier nonlinear phase look much similar to those found in the non-explosion models, with the power law being satisfied for $l \gtrsim 10$. In the later nonlinear phase, 
on the other hand, the power law is extended to much lower $l$'s. This is related to the late-time prominence of the $l=1$ mode as mentioned above. In fact, the amplitudes in 
the lower $l$ portion of the power spectrum are enhanced in $250 \le t \le 400$. The mechanism of this amplification is remaining to be understood but might be related with 
the volume-filling thermal convection advocated for the late stage of the convective instability in the supernova core by \citet{kifo}. 
As a result of this evolution of the spectrum, the shock tends to be ellipsoidal as the shock expands. 

\subsection{Neutrino Heating}

The SASI is supposed to be an important ingredient not only for the pulsar's proper motions and spins but also for the explosion mechanism itself. 
In this section, we look into the neutrino heating in the non-axisymmetric SASI. 

\begin{figure}[hbt]
\begin{picture}(0,450)
\put(10,240){\includegraphics[width=75mm]{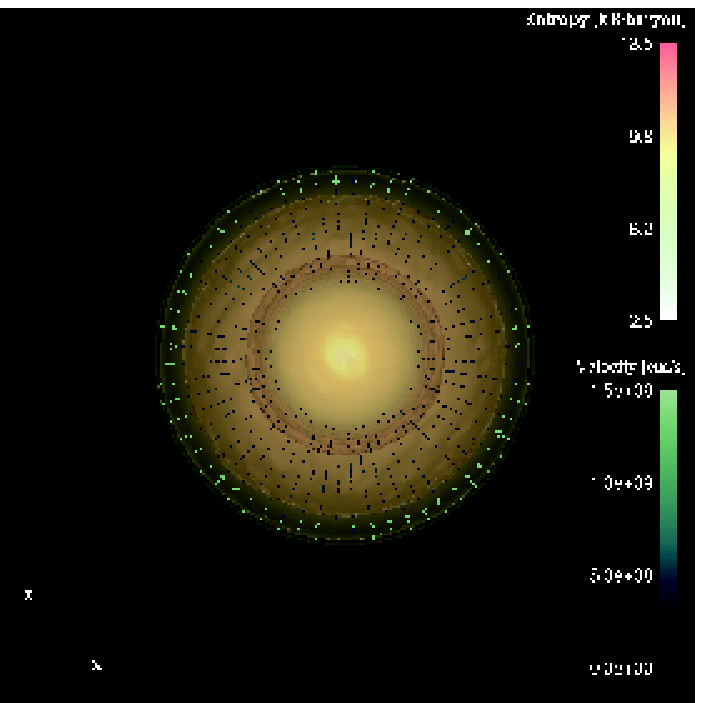}}
\put(10,230){\scalebox{1.1}{\footnotesize (a) $t=40$ ms}}
\put(235,240){\includegraphics[width=75mm]{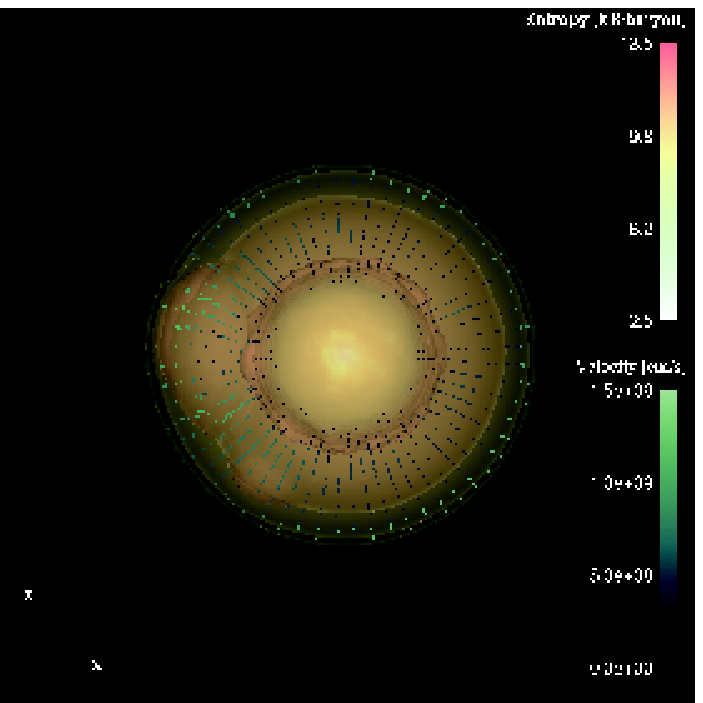}}
\put(235,230){\scalebox{1.1}{\footnotesize (b) $t=90$ ms}}
\put(10,10){\includegraphics[width=75mm]{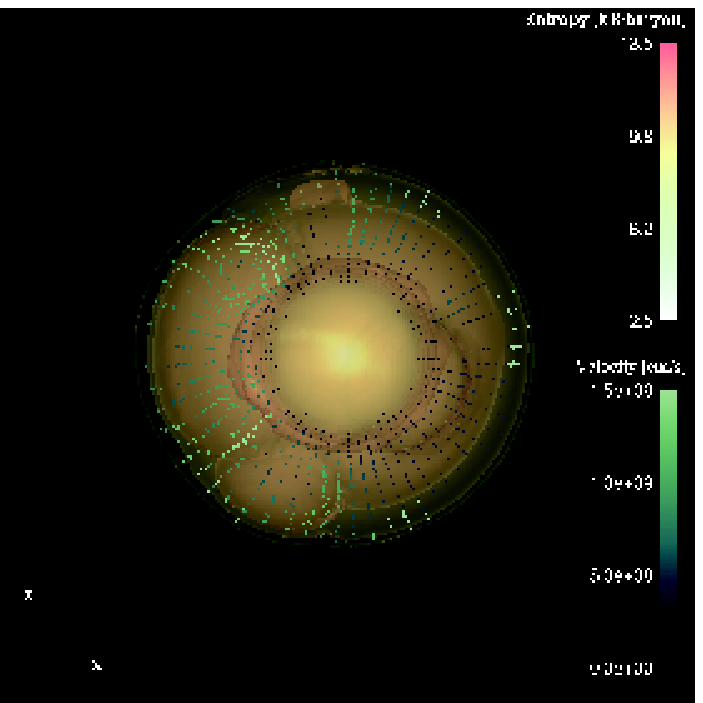}}
\put(10,0){\scalebox{1.1}{\footnotesize (c) $t=100$ ms}}
\put(235,10){\includegraphics[width=75mm]{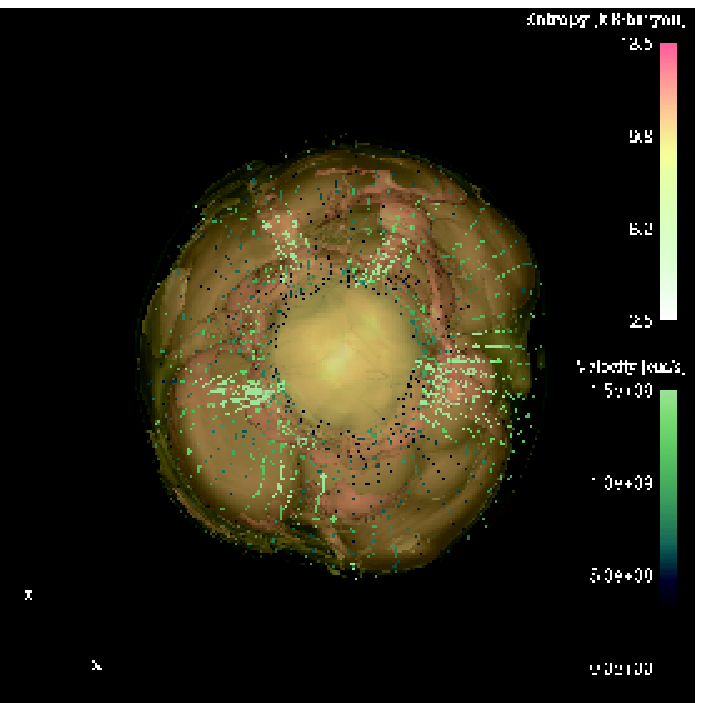}}
\put(235,0){\scalebox{1.1}{\footnotesize (d) $t=390$ ms}}
\end{picture}
\caption{The iso-entropy surfaces and velocity vectors in the meridian section for Model I\kern-.1emV.
The hemispheres ($0\le \phi \le \pi$) of eight iso-entropy surfaces  
are superimposed on one another with the outermost surface nearly corresponding to the shock front.}
\label{fig7}
\end{figure}

\begin{figure}[hbt]
\begin{picture}(0,450)
\put(10,240){\includegraphics[width=75mm]{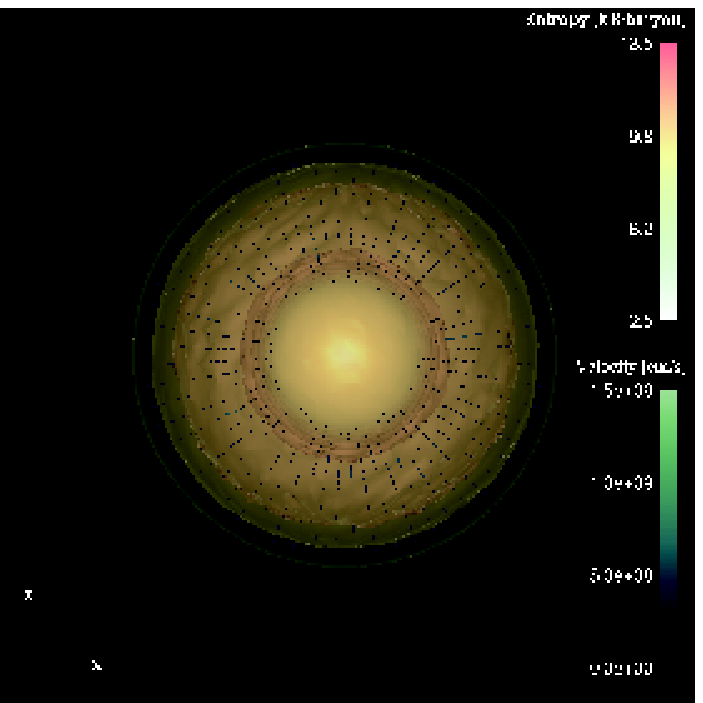}}
\put(10,230){\scalebox{1.1}{\footnotesize (a) $t=40$ ms}}
\put(235,240){\includegraphics[width=75mm]{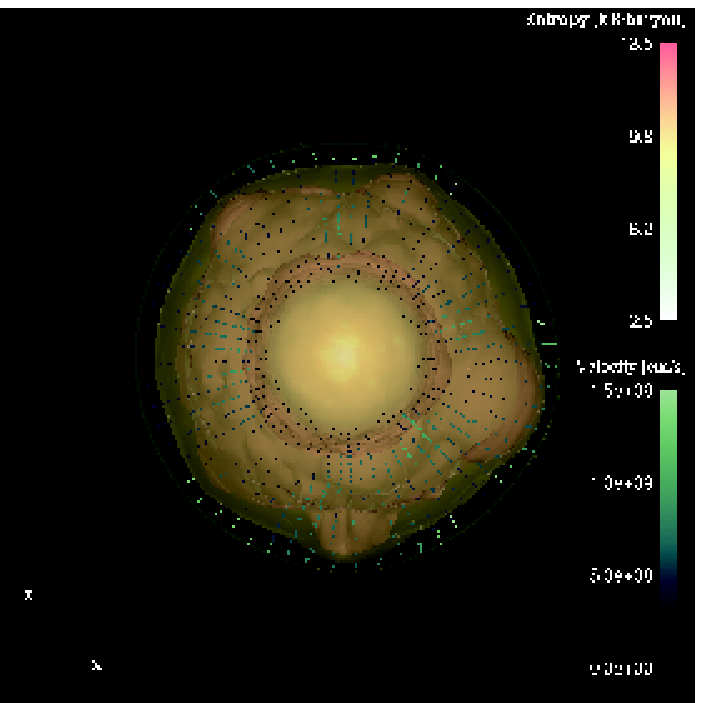}}
\put(235,230){\scalebox{1.1}{\footnotesize (b) $t=70$ ms}}
\put(10,10){\includegraphics[width=75mm]{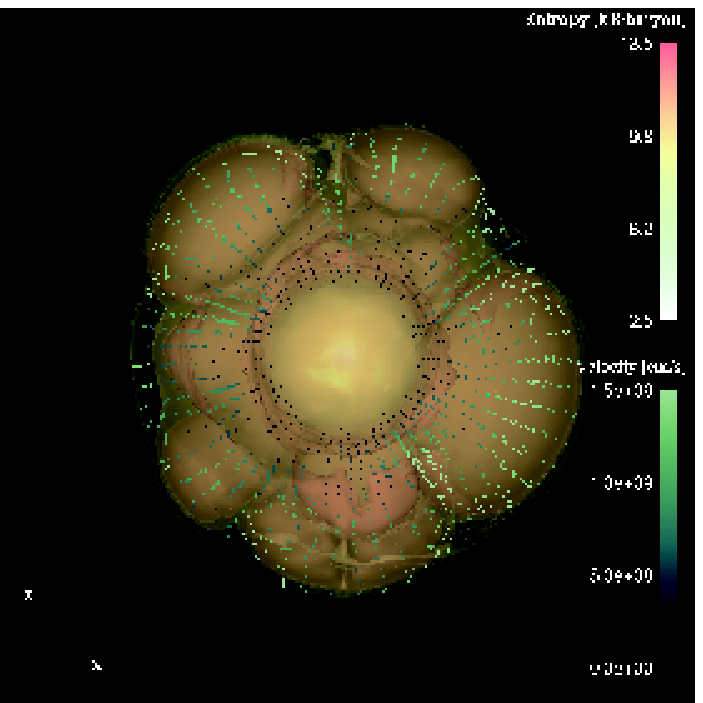}}
\put(10,0){\scalebox{1.1}{\footnotesize (c) $t=80$ ms}}
\put(235,10){\includegraphics[width=75mm]{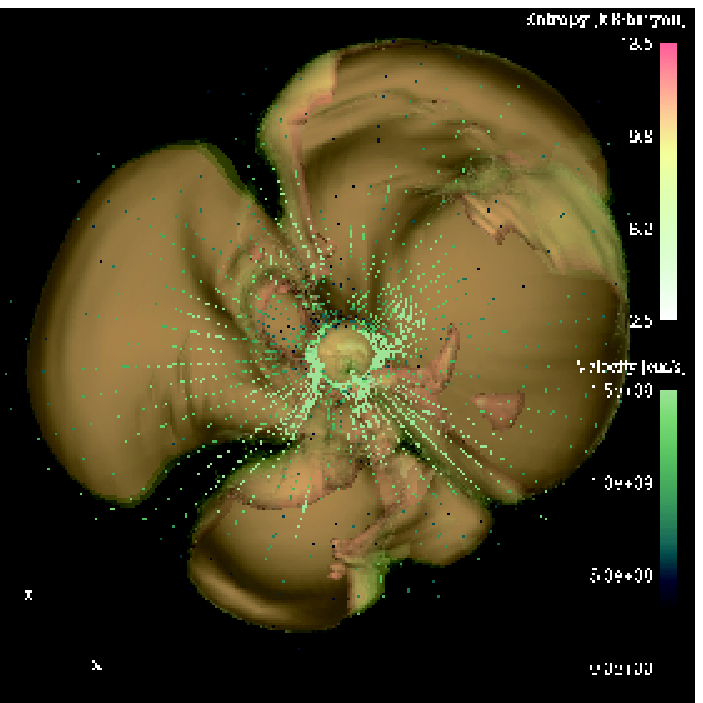}}
\put(235,0){\scalebox{1.1}{\footnotesize (d) $t=350$ ms}}
\end{picture}
\caption{The iso-entropy surfaces and velocity vectors in the meridian section for the explosion model (Model V\kern-.1emI). 
Note that the displayed region is 2.5 times larger for panel (d).}
\label{fig8}
\end{figure}

\begin{figure}[hbt]
\begin{picture}(0,200)
\put(10,10){\includegraphics[width=75mm]{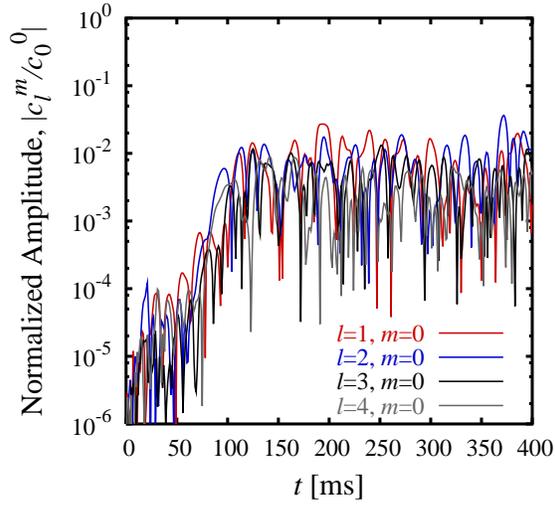}}
\put(10,0){\scalebox{1.1}{\footnotesize (a) Model I\kern-.1emV}}
\put(235,10){\includegraphics[width=75mm]{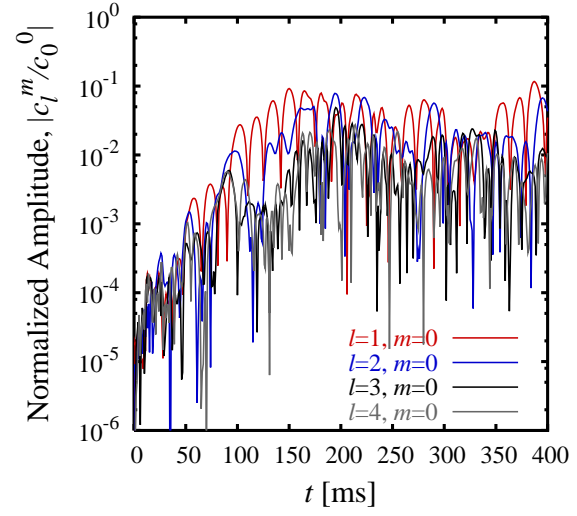}}
\put(235,0){\scalebox{1.1}{\footnotesize (b) Model V\kern-.1emI\kern-.1emI}}
\end{picture}
\caption{The time evolutions of the normalized amplitudes $|c^m_l/c^0_0|$ for Model I\kern-.1emV (non-axisymmetric) and 
V\kern-.1emI\kern-.1emI (axisymmetric). Note that the $m \neq 0$ modes also exist in the non-axisymmetric model but are not shown 
in this figure.}
\label{fig9}
\end{figure}

\begin{figure}[hbt]
\begin{picture}(0,570)
\put(   10,450){\includegraphics[width=75mm]{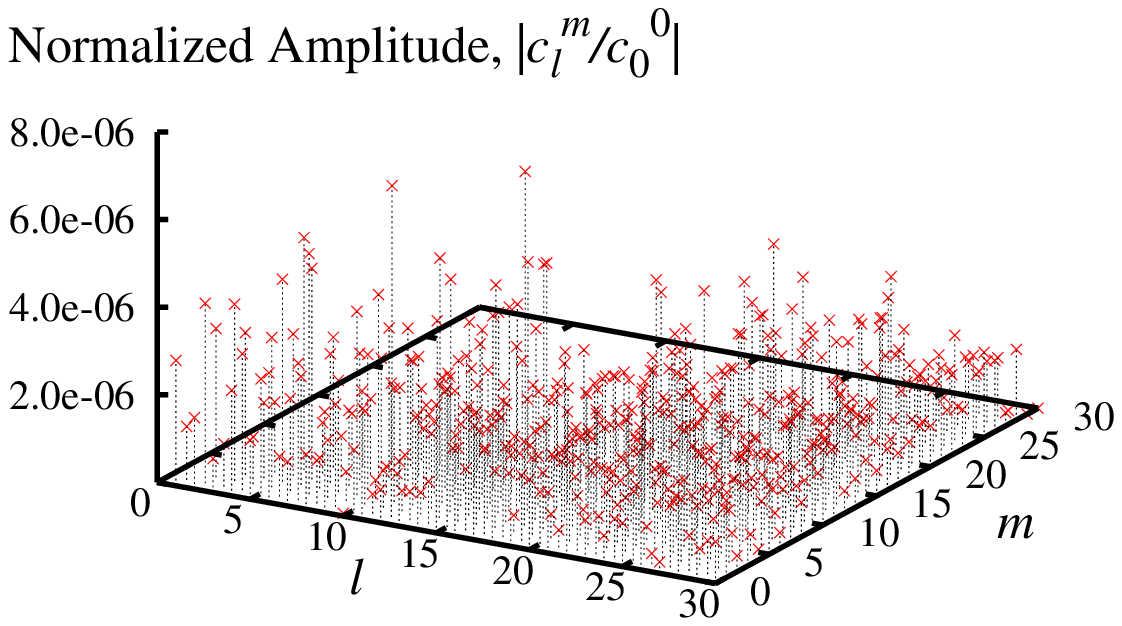}}
\put(   10,450){\scalebox{1.1}{\footnotesize (a) $t=1$ ms}}
\put(   10,300){\includegraphics[width=75mm]{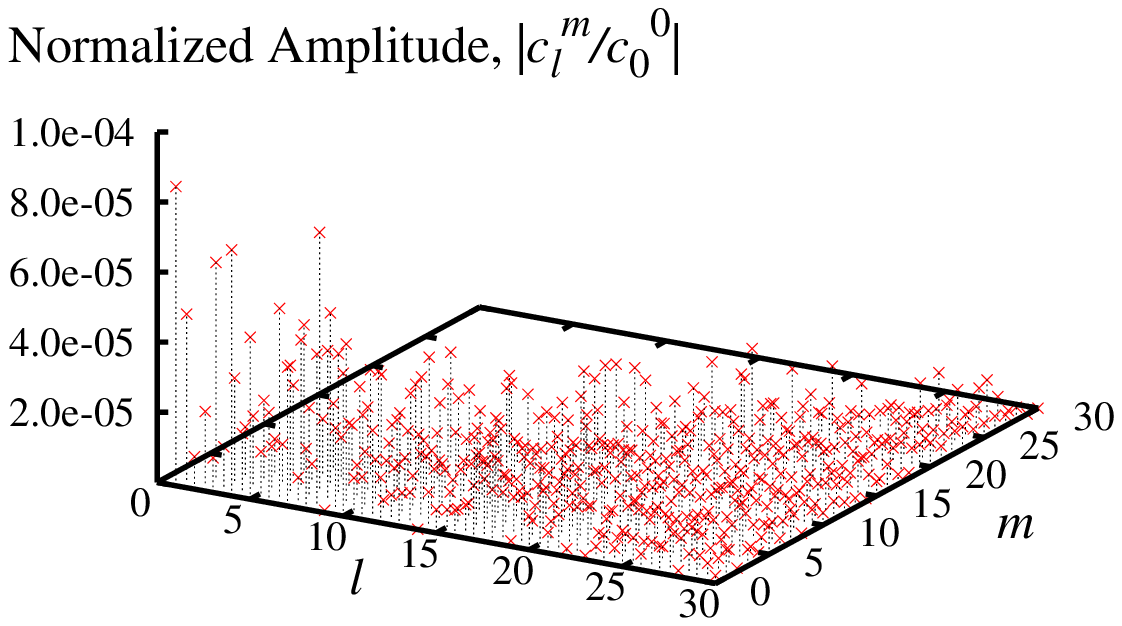}}
\put(   10,300){\scalebox{1.1}{\footnotesize (b) $t=30$ ms}}
\put(   10,150){\includegraphics[width=75mm]{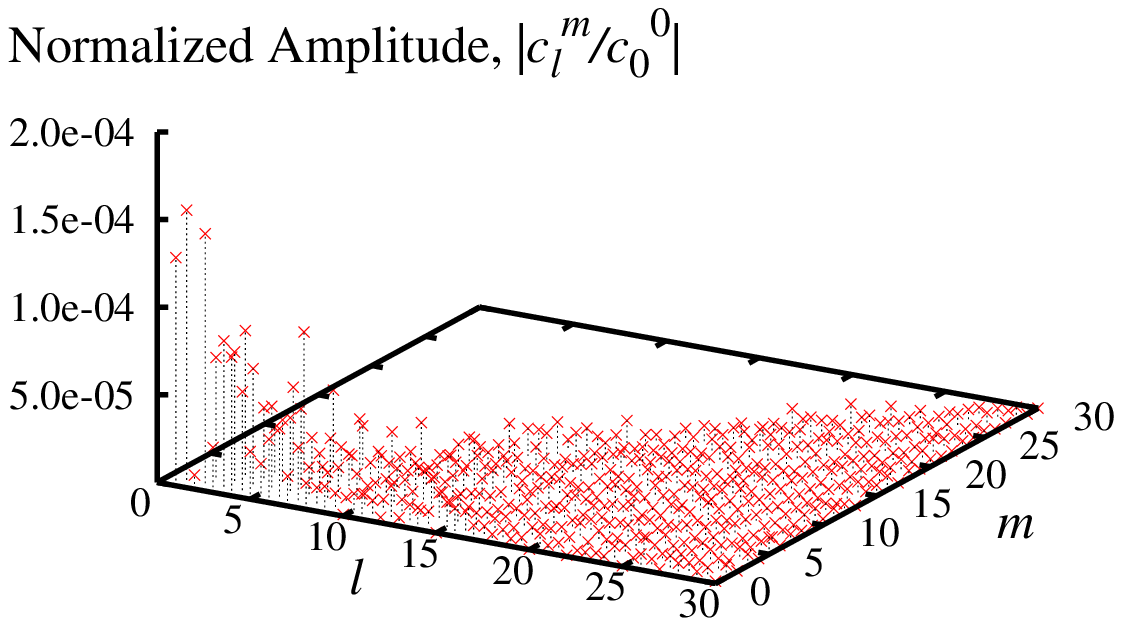}}
\put(   10,150){\scalebox{1.1}{\footnotesize (c) $t=60$ ms}}
\put(   10,     0){\includegraphics[width=75mm]{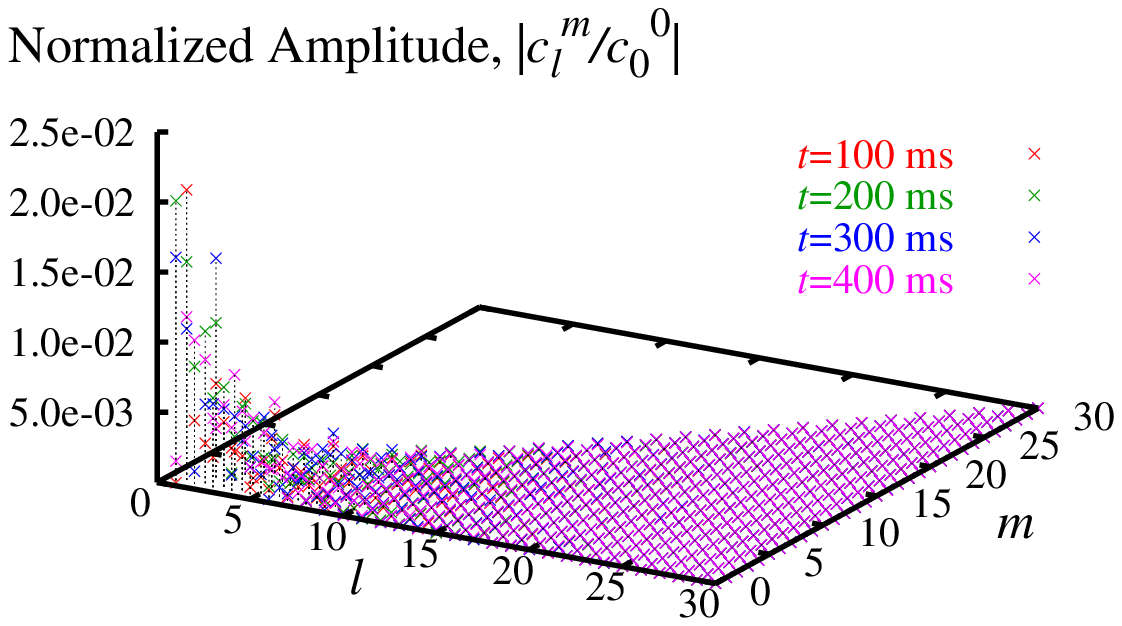}}
\put(   10,     0){\scalebox{1.1}{\footnotesize (d) $t=100-400$ ms}}
\put(235,450){\includegraphics[width=75mm]{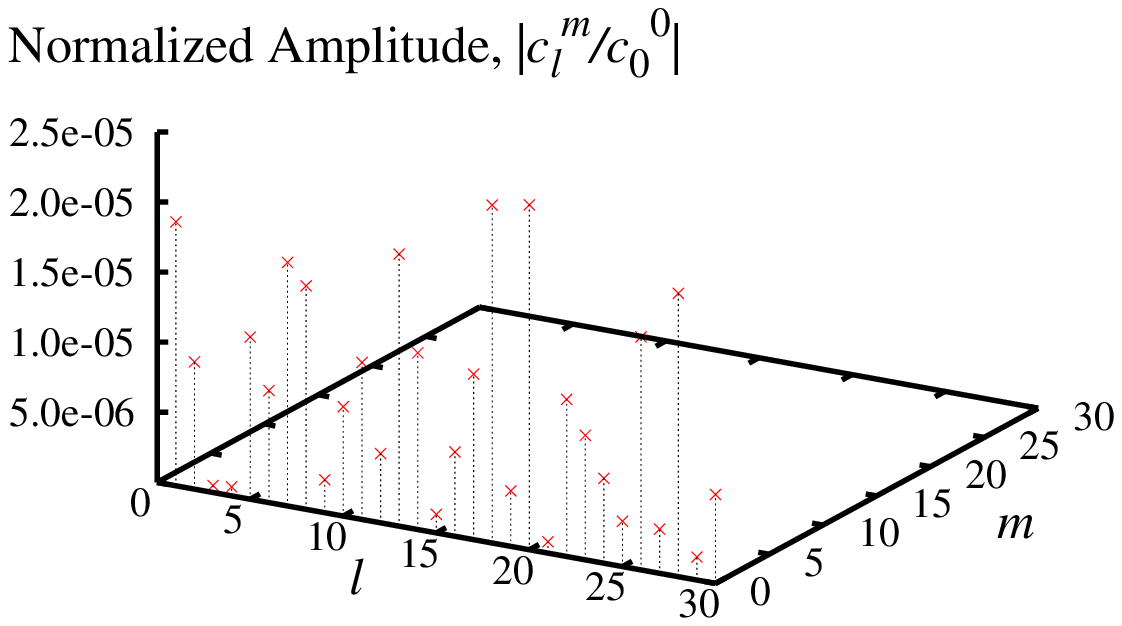}}
\put(235,450){\scalebox{1.1}{\footnotesize (e) $t=1$ ms}}
\put(235,300){\includegraphics[width=75mm]{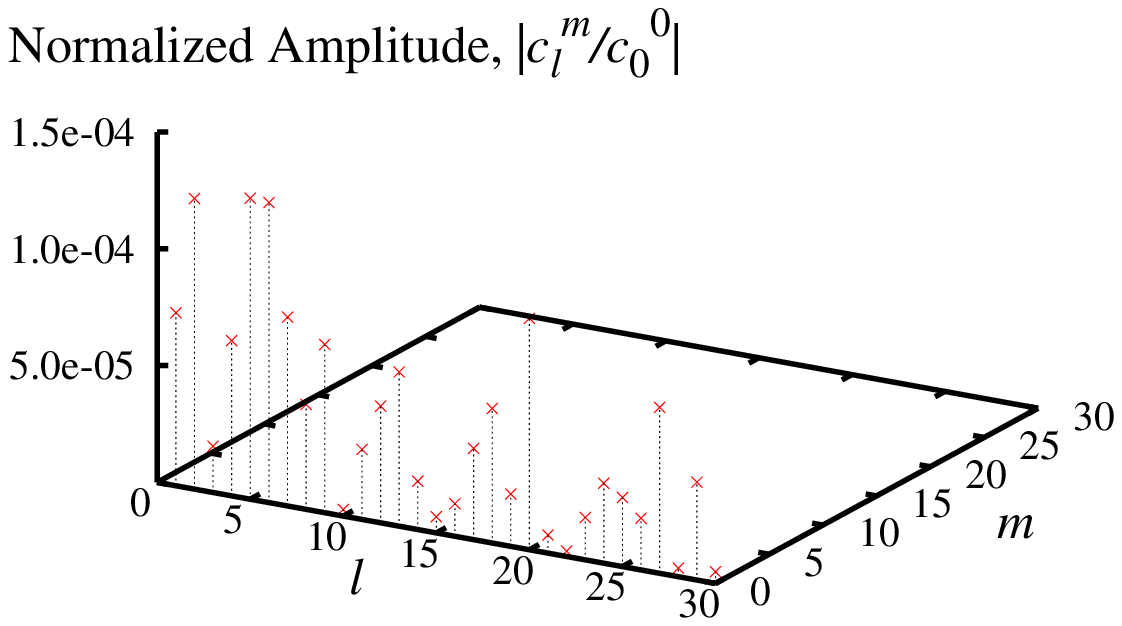}}
\put(235,300){\scalebox{1.1}{\footnotesize (f) $t=30$ ms}}
\put(235,150){\includegraphics[width=75mm]{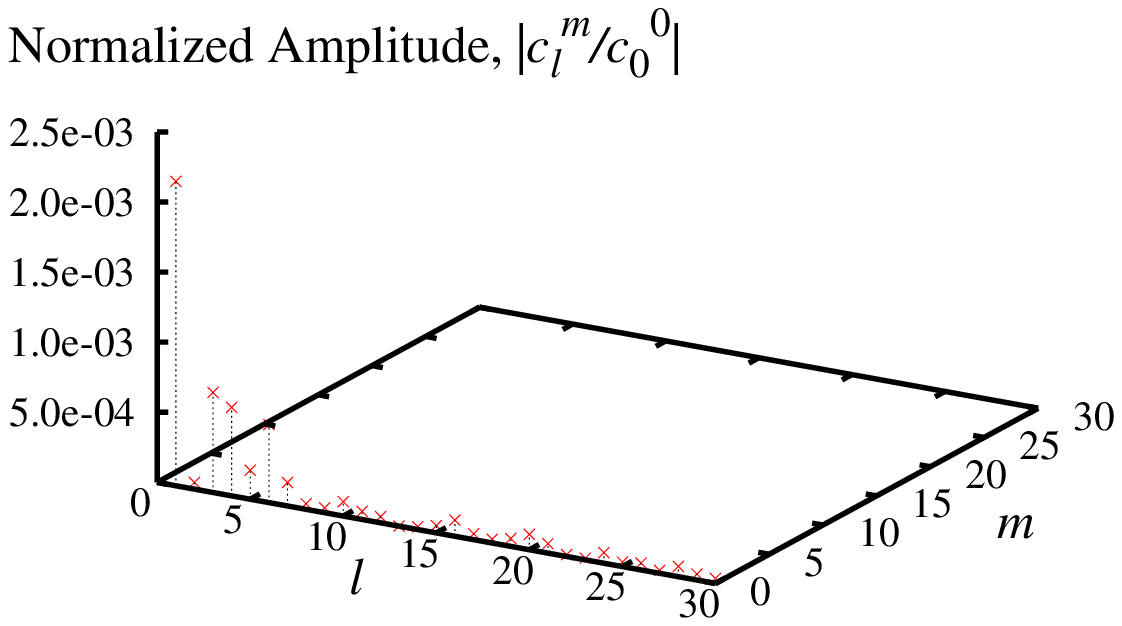}}
\put(235,150){\scalebox{1.1}{\footnotesize (g) $t=60$ ms}}
\put(235,     0){\includegraphics[width=75mm]{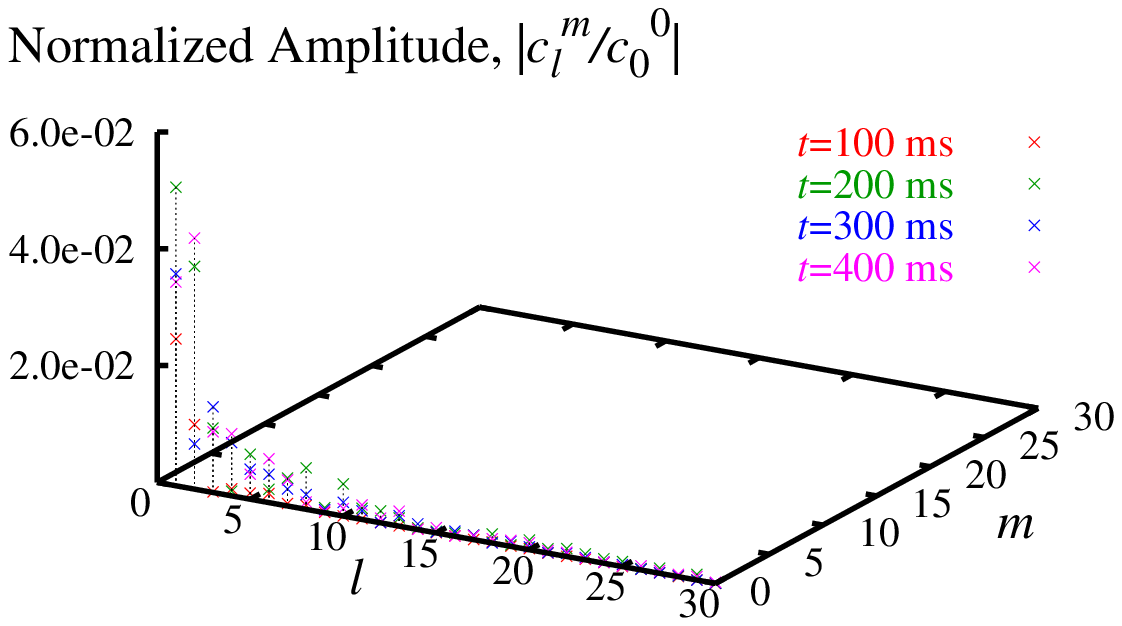}}
\put(235,     0){\scalebox{1.1}{\footnotesize (h) $t=100-400$ ms}}
\end{picture}
\caption{The normalized amplitudes $|c^m_l/c^0_0|$ for Model I\kern-.1emV (left panels) and Model V\kern-.1emI\kern-.1emI (right panels)
at different times. Note that the time-averaged values are plotted in panels (d) and (h).}
\label{fig10} 
\end{figure}

\begin{figure}[hbt]
\begin{picture}(0,200)
\put(10,0){\includegraphics[width=75mm]{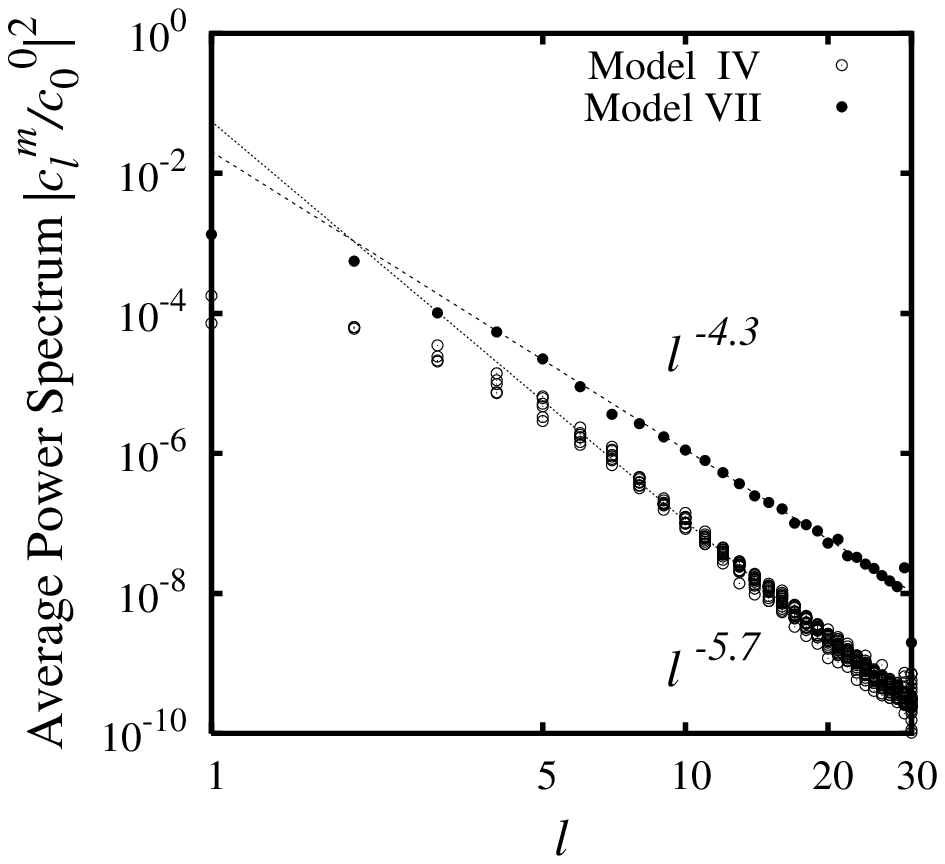}}
\put(10,0){\scalebox{1.1}{\footnotesize (a)}}
\put(235,0){\includegraphics[width=75mm]{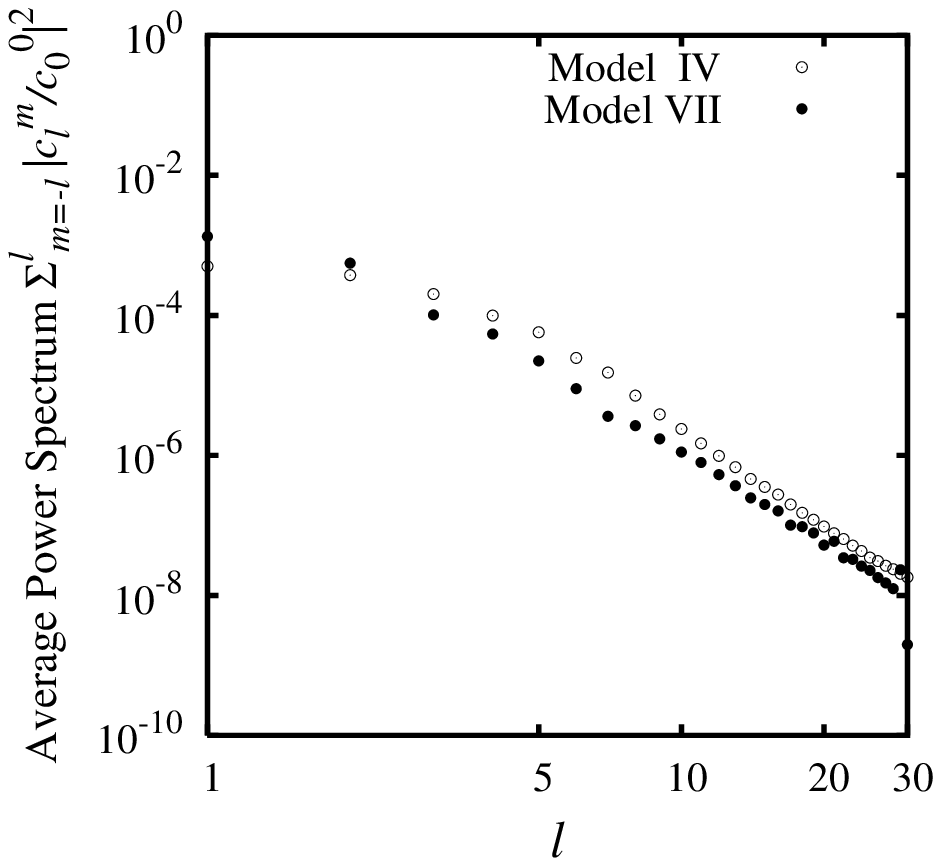}}
\put(235,0){\scalebox{1.1}{\footnotesize (b)}}
\end{picture}
\caption{The time-averaged power spectra for Model I\kern-.1emV and V\kern-.1emI\kern-.1emI. 
The average is taken over $t=150-400$ ms. In the left panel, modes with different $m$'s are plotted separately 
whereas they are summed up in the right panel.}
\label{fig11}
\end{figure}

\begin{figure}[hbt]
\begin{picture}(0,200)
\put(10,5){\includegraphics[width=75mm]{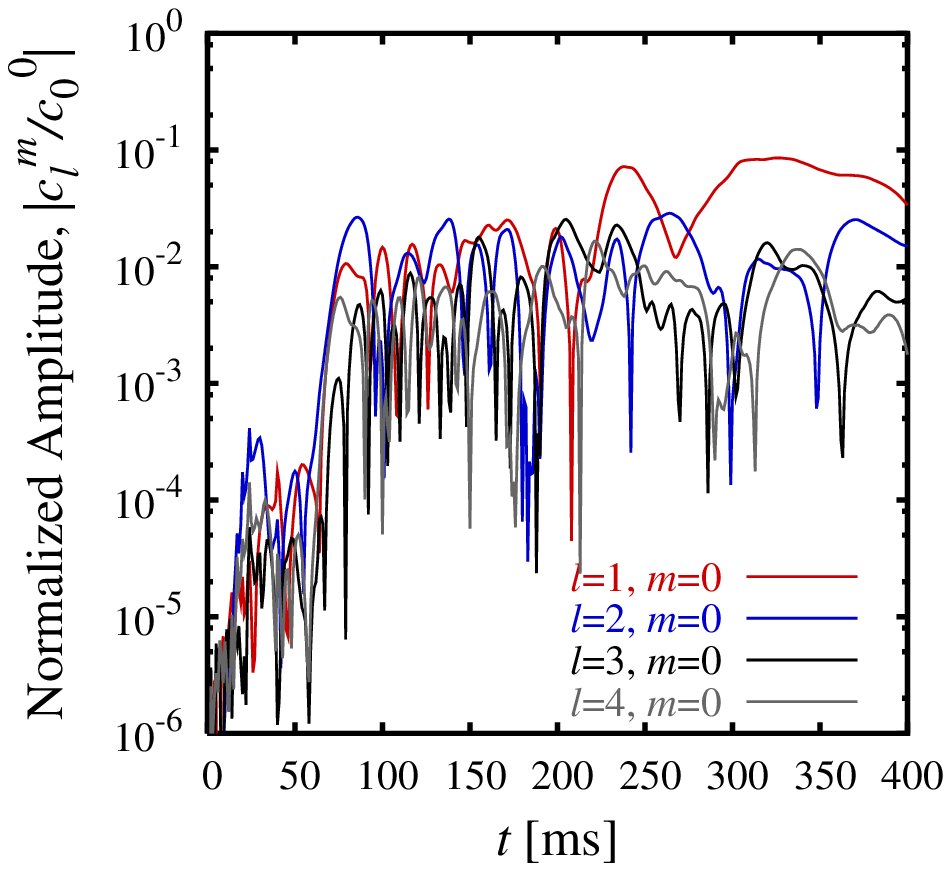}}
\put(10,0){\scalebox{1.1}{\footnotesize (a) Model V\kern-.1emI}}
\put(235,5){\includegraphics[width=75mm]{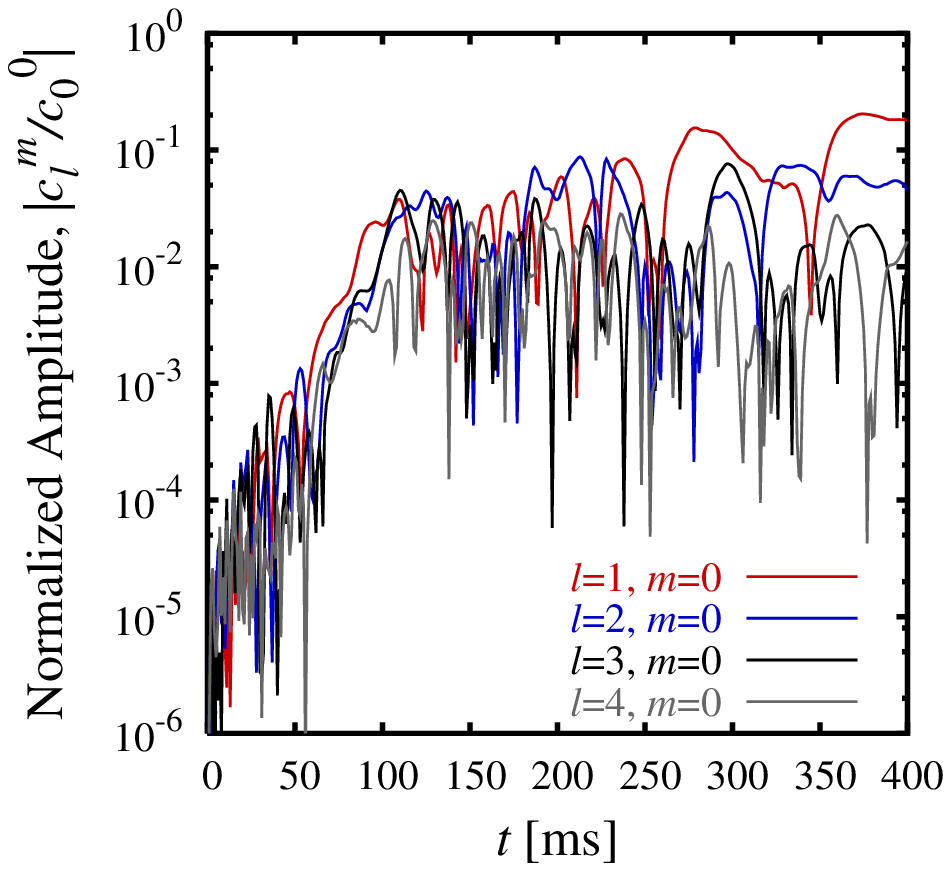}}
\put(235,0){\scalebox{1.1}{\footnotesize (b) Model V\kern-.1emI\kern-.1emI\kern-.1emI}}
\end{picture}
\caption{The time evolutions of the normalized  amplitudes $|c^m_l/c^0_0|$ for the explosion models (Model V\kern-.1emI (non-axisymmetric) in the left panel
and Model V\kern-.1emI\kern-.1emI\kern-.1emI (axisymmetric) in the right panel.}
\label{fig12}
\end{figure}

\begin{figure}[hbt]
\begin{picture}(0,200)
\put(10,0){\includegraphics[width=75mm]{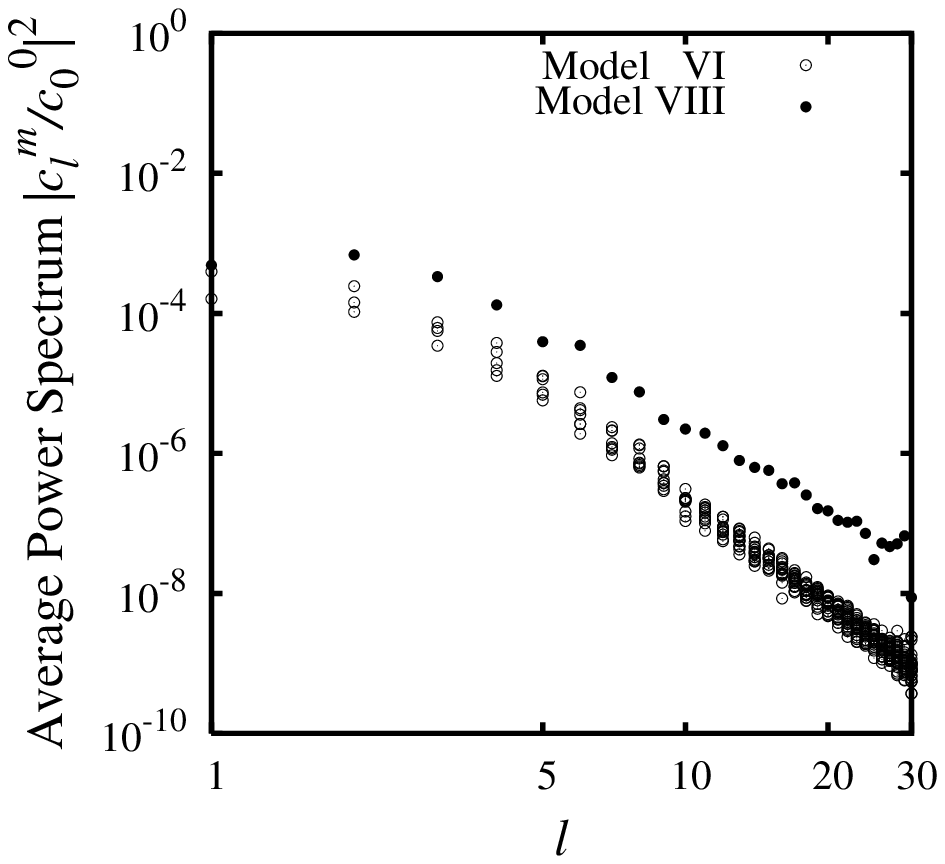}}
\put(10,0){\scalebox{1.1}{\footnotesize (a) $t=100-200$ ms}}
\put(235,0){\includegraphics[width=75mm]{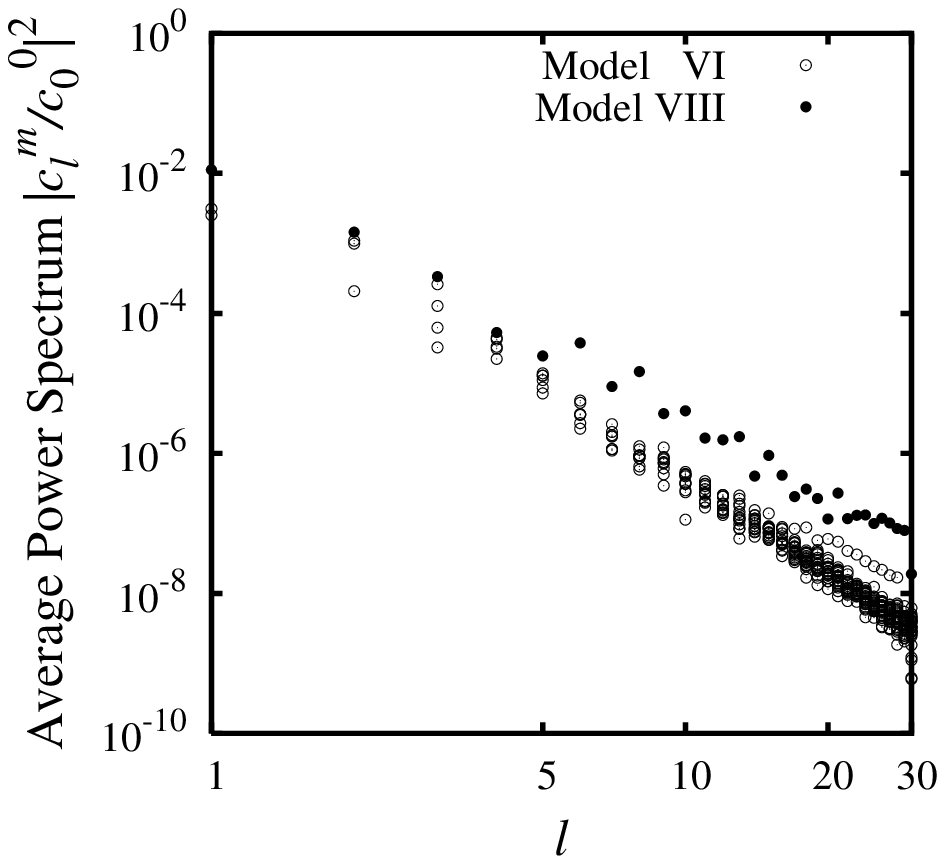}}
\put(235,0){\scalebox{1.1}{\footnotesize (b) $t=250-400$ ms}}
\end{picture}
\caption{The time-averaged power spectra for the explosion models, Model V\kern-.1emI (non-axisymmetric) and 
V\kern-.1emI\kern-.1emI\kern-.1emI (axisymmetric). The averages are taken (a) for $t=100-200$ ms and (b) for $t=250-400$ ms, respectively.}
\label{fig13}
\end{figure}

\begin{figure}[hbt]
\begin{picture}(0,600)
\put(30,410){\includegraphics[width=65mm]{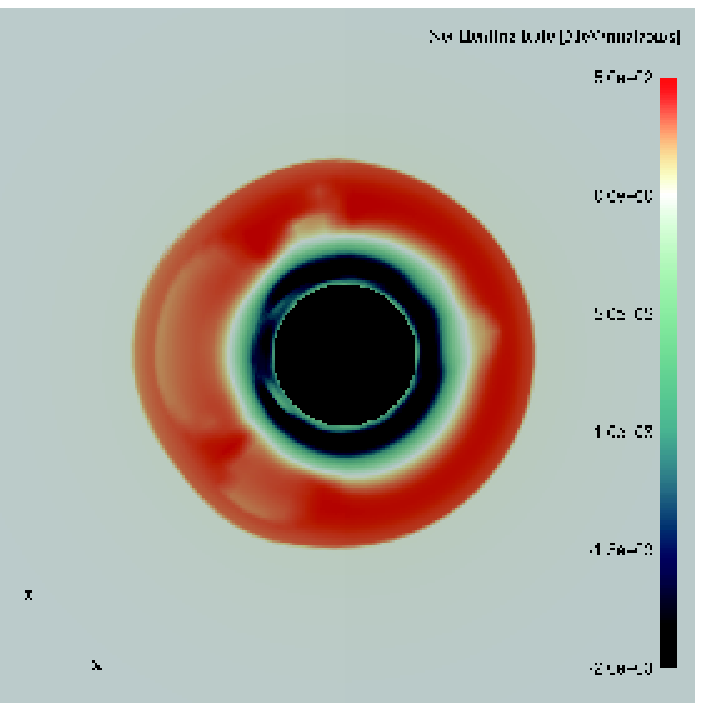}}
\put(30,400){\scalebox{1.1}{\footnotesize (a) Model I\kern-.1emV, $t=100$ ms}}
\put(235,410){\includegraphics[width=65mm]{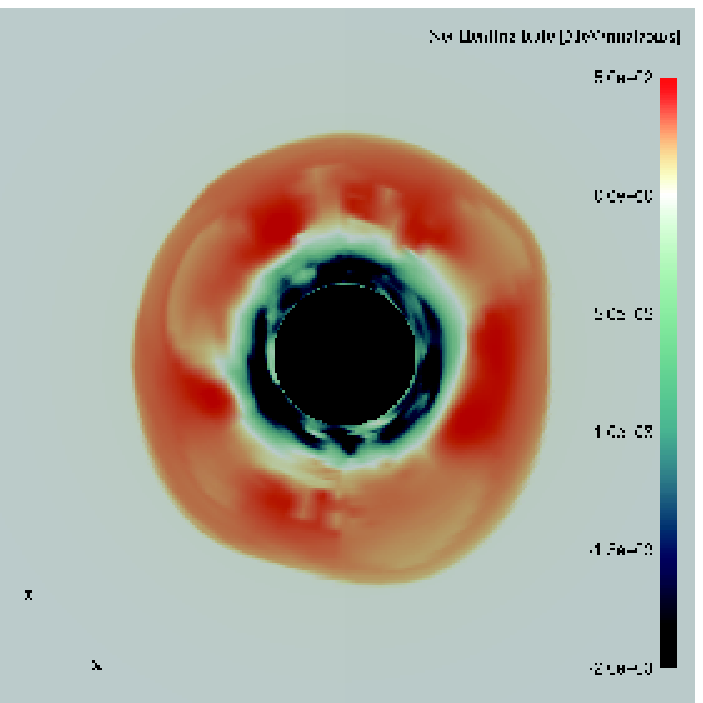}}
\put(235,400){\scalebox{1.1}{\footnotesize (b) Model I\kern-.1emV, $t=390$ ms}}
\put(30,210){\includegraphics[width=65mm]{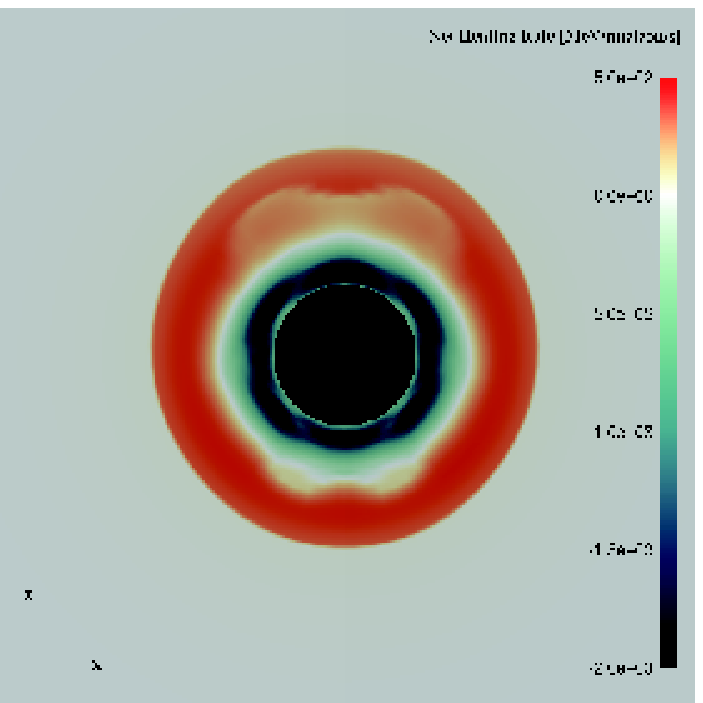}}
\put(30,200){\scalebox{1.1}{\footnotesize (c) ModelV\kern-.1emI\kern-.1emI, $t=100$ ms}}
\put(235,210){\includegraphics[width=65mm]{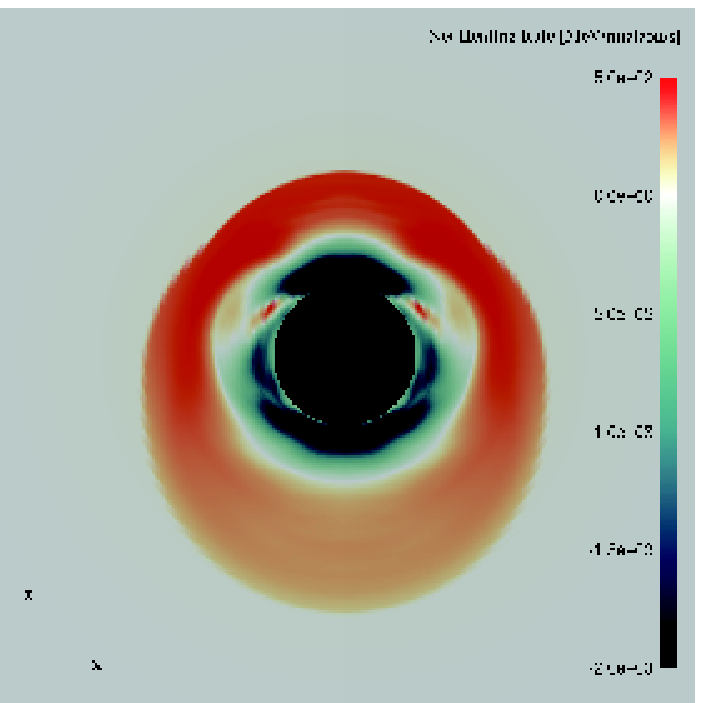}}
\put(235,200){\scalebox{1.1}{\footnotesize (d) V\kern-.1emI\kern-.1emI, $t=390$ ms}}
\put(30,10){\includegraphics[width=65mm]{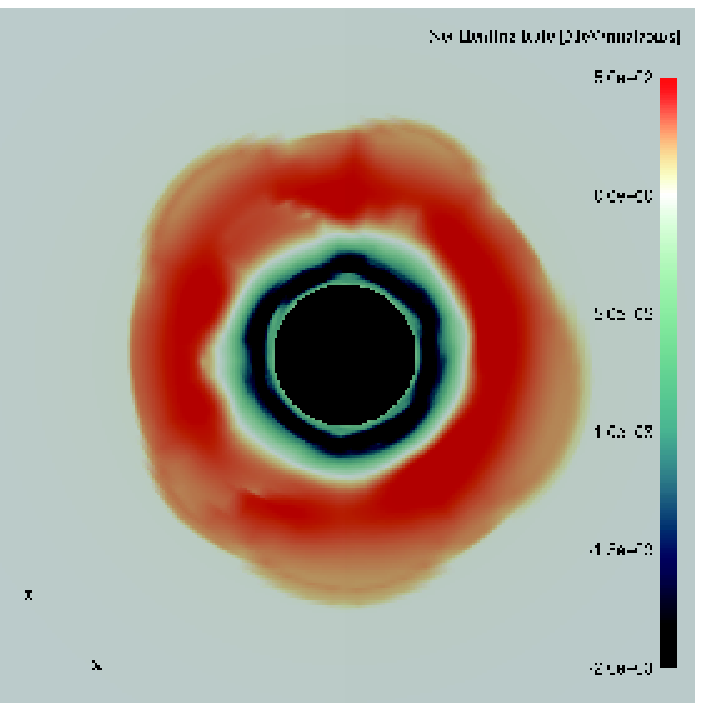}}
\put(30,0){\scalebox{1.1}{\footnotesize (e) Model V\kern-.1emI, $t= 80$ ms}}
\put(235,10){\includegraphics[width=65mm]{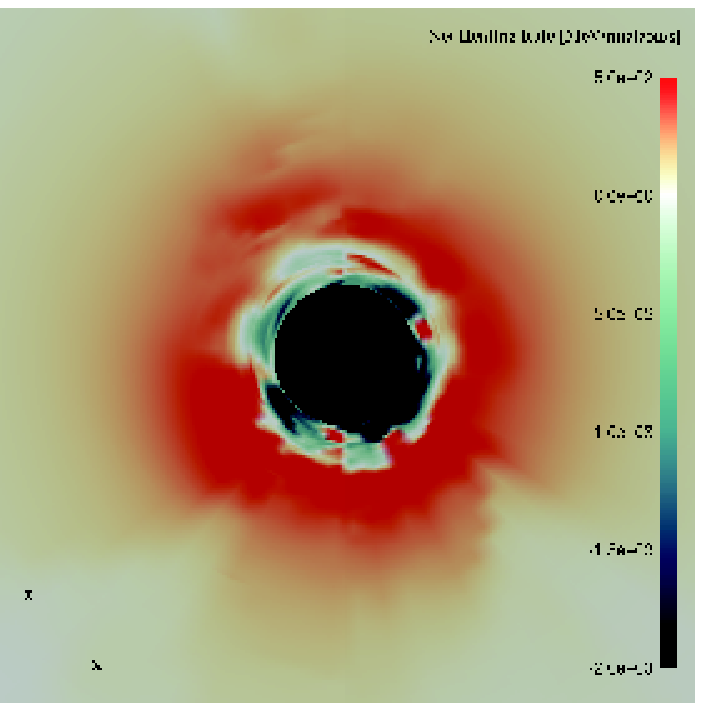}}
\put(235,0){\scalebox{1.1}{\footnotesize (f) Model V\kern-.1emI, $t= 350$ ms}}
\end{picture}
\caption{The color contours of the net heating rate in the meridian section for 
Model I\kern-.1emV  (top panels), Model V\kern-.1emI\kern-.1emI (middle panels) and Model V\kern-.1emI (bottom panels)
at different times.}
\label{fig14}
\end{figure}

We show in Figure~\ref{fig14} the color contours of the net heating rate in the meridian section 
for the non-explosion model with $L_\nu=6.0 \times 10^{52}$ ergs s$^{-1}$ (non-axisymmetric Model I\kern-.1emV and 
axisymmetric Model V\kern-.1emI\kern-.1emI) as well as for the non-axisymmetric explosion model with $L_\nu=6.8 \times 10^{52}$ ergs s$^{-1}$ (Model V\kern-.1emI ). 
In the early phase, the cooling region with negative net heating rates exists around the proto neutron star while the heating region extends
over the cooling region up to the stalled shock wave in all the cases.
As the time passes and the SASI grows, a pocket of regions with positive but relatively low net heating rates appear.
These regions correspond to the high entropy blobs (high entropy rings for the axisymmetric case), where the circulating flow exists observed 
in Figure~\ref{fig7} (b), (c), and (d). 
Since the neutrino emission in these regions is more efficient 
than the surroundings, the net heating rate is rather low. 

Although the critical neutrino luminosties are not much different between the non-axisymmetric and axisymmetric cases, 
the spatial distributions of neutrino heating are different. In the axisymmetric case the shock wave oscillates up and down whereas it moves in all 
directions in the non-axisymmetric case. The oscillation amplitudes are larger in the former than in the latter in general as repeated mentioned. 
Reflecting this difference in the shock motions, the neutrino heating is enhanced in mainly in the polar regions in the axisymmetric case
while in the non-axisymmetric case the heating rate is affected by SASI chiefly around the high entropy blobs. These are clearly seen 
in Figure~\ref{fig14} (see also Figure~\ref{fig7}).

As described in section~\ref{basic}, the generation of the high entropy blobs is started around the end of the linear phase and continues during the nonlinear phase. 
Although the blobs repeat expansions and contractions, being sometimes merged or splitted, the turbulent motions together
with the neutrino heating are finally settled to a quasi-steady state in the non-explosion model. 
For the explosion model, on the other hand, the number and volume of high entropy blobs are
increased much more quickly and, as a result, the heating region prevails, pushing the shock wave outwards and narrowing the cooling region near the proto neutron star.

\begin{figure}[hbt]
\begin{picture}(0,200)
\put(125,0){\includegraphics[width=75mm]{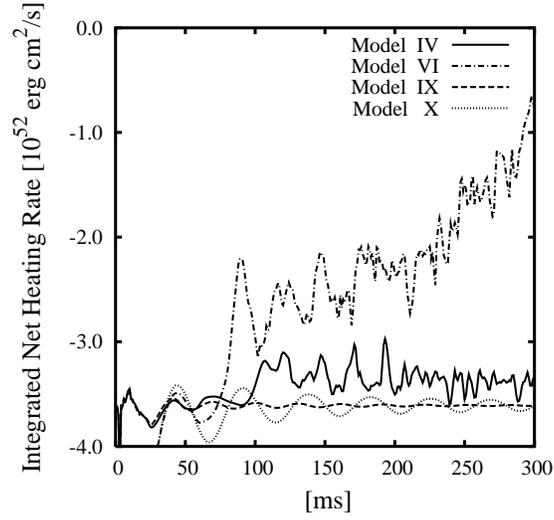}}
\end{picture}
\caption{The time variations of the net heating rate integrated over the region inside the shock wave. 
Models I\kern-.1emV  and V\kern-.1emI are compared with the spherically symmetric counterparts, Models I\kern-.1emX and X.}
\label{fig15}
\end{figure}

Figure~\ref{fig15} shows the time evolutions of the net heating rates integrated over the region inside the shock wave.
For the non-explosion model (Model I\kern-.1emV, solid line), the volume-integrated heating rate grows until $t \sim 150$ms but is saturated thereafter,
whereas it increases continuously for the explosion model (Model V\kern-.1emI, dashed-dotted line). These behaviors of the volume-integrated heating rate are 
in accord with the time evolutions of the average shock radius as shown in Figure~\ref{fig6} (a). It is clear from the figure that the heating rates are 
mainly affected by SASI during the nonlinear phase. For comparison, the spherically symmetric counterparts, Models I\kern-.1emX with $L_\nu=6.0 \times 10^{52}$ ergs s$^{-1}$ and 
X with $L_\nu=6.8 \times 10^{52}$ ergs s$^{-1}$ are also presented as dashed and dotted lines, respectively. Note that the last model does not lead to the shock 
revival even with this high neutrino luminosity. 
It is found that the SASI enhances the neutrino heating both for the explosion and non-explosion models.

\section{Conclusion \label{sec4}}

In this paper we have studied the non-axisymmetric SASI by 3D hydrodynamical simulations, taking into account the realistic EOS and neutrino-heating and cooling. We have added
various non-axisymmetric perturbations to spherically symmetric steady flows that accrete through a standing shock wave onto a proto-neutron star, being irradiated by neutrinos 
emitted from the proto-neutron star. The mode analysis has been done for the deformation of the shock surface by the spherical harmonics expansion. After confirming that our 3D 
code is able to reproduce for the axisymmetric perturbations the previous results on 2D SASI that we obtained in \citet{ohnishi_1}, we have done genuinely 3D simulations and found 
the followings.

First, the model of the initial perturbation with $(l, |m|)=(1,0)$ and $(1,1)$ have demonstrated that the non-axisymmetric SASI proceeds much in the same manner as the
axisymmetric SASI: the linear phase, in which the initial perturbation grows exponentially, lasts for about $100$ms and is followed by the nonlinear phase, in which various modes are
produced by nonlinear mode couplings and their amplitudes are saturated. It has been found that the critical neutrino luminosity, above which the shock revival occurs, is not much different
between 2D and 3D. For the neutrino luminosities lower than the critical value, the SASI is settled to a quasi-steady turbulence. We have found that the
saturation level of each mode in the non-axisymmetric SASI is lower in general than that of the axisymmetric counterpart. This is mainly due to the fact that the number of the modes
contributing to the turbulence is larger for the non-axisymmetric case. The sequence of the mode generation, on the other hand, strongly suggests that the nonlinear mode coupling is
chiefly of quadratic nature. 

Second, the simulations with the random multi-mode perturbations being imposed initially have shown that the dynamics in the linear phase is essentially a superimposition of 
those of single-modes. Toward the end of the linear phase, high entropy blobs are generated continuously and grow, starting the nonlinear phase.
We have observed that these blobs repeat expansions and contrations, being merged and splitted from time to time.
In the non-explosion models, these nonlinear dynamics lead to the saturation of mode amplitudes and the quasi-steady
turbulence. For the explosion models, on the other hand, the production of blobs proceeds much more quickly, followed by the oligarchic evolution, with a relatively small number of great 
blobs absorbing smaller ones, and as a result the shock radius increases almost monotonically. The spectral analysis has clearly demonstrated that low $l$ modes are predominant 
in the nonlinear phase just as in the axisymmetric case. We have also shown that the equi-partition is nearly established among different $m$ modes in the quasi-steady turbulence and
that the spectrum summed over $m$ in the non-axisymmetric case is quite similar to the axisymmetric counterpart. This implies that the larger number of modes in the non-axisymmetric 
case is the main reason why the amplitude of each mode is smaller in 3D than in 2D. The power spectrum is approximated by the power law for $l \gtrsim 10$. Although the slope is 
slightly steeper for the non-axisymmetric models, whether the difference is significant or not is unknown at present.    

We have seen in the explosion models, on the other hand, that the oscillation period of each mode becomes longer in the late nonlinear phase as the shock radius gets larger and 
the advection-acoustic cycle becomes longer. What is more interesting is the fact that in this late phase the dominance of low $l$ modes becomes even more remarkable. Although 
this may be related with the volume-filling thermal convection, the mechanism is yet to be revealed. This spectral evolution leads to the global deformation of the shock surface to 
an ellipsoidal shape, whose major axis is not necessarily aligned with the symmetry axis, however.

Finally, we have presented the neutrino heating in the 3D SASI. It has been seen that the volume-integrated heating rate is affected mainly in the nonlinear phase.
The comparison with the spherically symmetric counterparts has confirmed that the SASI enhances the neutrino heating also in the non-axisymmetric case.
Although the critical neutrino luminosity in the non-axisymmetric SASI is not much changed from that for the axisymmetric case, the spatial distribution of neutrino heating 
is different in the non-linear phase.
Relatively narrow regions surrounding high entropy blobs are efficiently heated for the non-axisymmetric case,
while wider regions near the symmetry axis are heated strongly in accord with the sloshing motion of shock wave along the symmetry axis for
the axisymmetric case. For the non-explosion models, the high entropy blobs produced by neutrino heatings occupy the inside of shock wave, repeating expansion and contraction 
and being splitted and merged intermittently, but the flow is finally settled to a quasi-static state.
For the explosion models, on the other hand, the high entropy blobs are generated much more quickly and extend the heating region, pushing the shock outwards 
and narrowing the cooling region near the proto-neutron star. 

In the present study we have not observed a persistent segregation of angular momentum in the accretion flow such as found by \citet{blondin_nat, blondin_shaw}) although 
instantaneous spiral flows are frequently seen. As discussed above, the equi-partition is nearly established among different $m$ modes in our models. It should be emphasized 
here again, however, that we have added the initial perturbations only to the radial velocity in this study and, as a result, the modes with $m=\pm1$ are equally contributing. 
We will defer the analysis of the models with the initial non-axisymmetric perturbations being added also to the azimuthal component of velocity to our forthcoming paper~\citep{iwa07}, 
in which we will also discuss a possible correlation between the kick velocity and spin of neutron stars if they are produced by the 3D SASI indeed. A lot of issues on SASI are still 
remaining to be studied. In particular, the influences of rotation and magnetic field are among the top priorities and will be addressed elsewhere in the near future. 

\acknowledgements{
One of the authors (WI) expresses her sincere gratitude to assoc. prof. Kazuyuki Ueno, assistant prof. Michiko Furudate, and the member of Sawada and Ueno laboratory for continuing encouragements and advices.
She would like to thank also Yoshitaka Nakashima of Tohoku University for his advices on the visuallization by AVS.
KK expresses his thanks to Katsuhiko Sato for his 
continuous encouragements. Numerical cumputations were performed on the Altix3700Bx2 at the Institute of Fluid Science, Tohoku University as well as on VPP5000 and the general common use computer system at the center for Computational Astrophysics, CfCA, the National Astronomical Observatory of Japan.
This study was partially supported by Program for Improvement of Research Environment for Young Researchers from Special Coordination Funds for Promoting Science and 
Technology (SCF), the Grants-in-Aid for the Scientific Research (No.S19104006, No.S14102004, No.14079202, No.14740166) and Grant-in-Aid for the 21st century COE program 
``Holistic Research and Education Center for Physics of Self-organizing Systems'' of Waseda University by the Ministry of Education, Culture, Sports, Science, and Technology (MEXT) of Japan.}

\appendix
\begin{center}
\section{Tensor Artificial Viscosity\label{visc}}
\end{center}
We present the tensor-type artificial viscosity, $\mathbf{Q}$, that we use in this paper. It is a direct extention to 3D of the one employed by \citet{stone} for 2D simulations.
Following the notations in \citet{stone}, $\mathbf{Q}$ is written as follows:
\begin{equation}
\mathbf{Q} = \begin{cases}  
                               l^2 \rho \nabla \cdot \bm{v}
                                      \left[\nabla \bm{v} -\frac{1}{3}\left(\nabla \cdot \bm{v} \right) \mathbf{e} \right] 
					                     & \mathrm{if} \ \ \ \nabla \cdot \bm{v} < 0 \\
					                    0 & \mathrm{otherwise}
					        \end{cases},
\end{equation}
where $l$ denotes a constant with a dimension of length, $\nabla \bm{v} = (\bm{v}_{i,j}+\bm{v}_{j,i})/2$ is the symmetrized velocity-gradient tensor, $\mathbf{e}$ is the unit tensor.
Dropping the off-diagonal components, we utilize only the diagonal components of $\mathbf{Q}$ in this study, which are written in the finite difference form as 
\begin{equation}
\begin{split}
\textstyle
Q11_{\ i,j,k} = l^2_{i,j,k} d_{i,j,k} \left(\nabla \cdot \bm{v}\right)_{i,j,k} 
                         \left[\left(\nabla \bm{v}_{(11)}\right)_{i,j,k} - 
					                \frac{1}{3}\left(\nabla \cdot \bm{v} \right)_{i,j,k} \right], \\
\textstyle
Q22_{\ i,j,k} = l^2_{i,j,k} d_{i,j,k} \left(\nabla \cdot \bm{v}\right)_{i,j,k} 
                         \left[\left(\nabla \bm{v}_{(22)}\right)_{i,j,k} - 
					                \frac{1}{3}\left(\nabla \cdot \bm{v} \right)_{i,j,k} \right],\\
\textstyle									
Q33_{\ i,j,k} = l^2_{i,j,k} d_{i,j,k} \left(\nabla \cdot \bm{v}\right)_{i,j,k} 
                         \left[\left(\nabla \bm{v}_{(33)}\right)_{i,j,k} - 
					                \frac{1}{3}\left(\nabla \cdot \bm{v} \right)_{i,j,k} \right],
\end{split}
\end{equation}
 where $d_{i,j,k}$ stands for the density at the site specified by $(i, j, k)$ and $l_{i,j,k}$ is written as
 \begin{equation}
 l_{i,j,k}=\mathrm{max} (C_2 \ dx1a_i, C_2 \ dx2a_j, C_2 \ dx3a_k).
 \end{equation}
 Here $C_2$ is a dimensionless constant controlling the number of grid points, over which a shock is spread, 
 and $dx1a_i$, $dx2a_j$, and $dx3a_k$ are the grid widths of the ``a-mesh" at the $i$-, $j$-, and $k$-th grid points in the $r$, $\theta$, and $\phi$ directions, respectively.
 The ``a-mesh" and ``b-mesh" are distinguished in ZEUS-MP/2 and are defined on the cell-edge and cell-center, respectively (see \citet{stone} for more details). 
 
The artificial viscous stress $-\nabla \cdot \mathbf{Q}$ and artificial dissipation $-\mathbf{Q}:\nabla \bm{v}$ in the momentum equation (\ref{eq:momentum}) and 
energy equation (\ref{eq:energy}) are given as follows,
 \begin{equation}
 \begin{split}
 \left( \nabla \cdot \mathbf{Q} \right)_{(1)} &= \frac{1}{g^2_2 g_{31}}\frac{\partial}{\partial x_1}
                                                                                             \left(g^2_2 g_{31} Q_{11} \right), \\
\left( \nabla \cdot \mathbf{Q} \right)_{(2)} &= \frac{1}{g_2 g^2_{32}}\frac{\partial}{\partial x_2}
                                                                                             \left(g^2_{32} Q_{22}\right) + \frac{Q_{11}}{g_2 g_{32}}
                                                                                             \frac{\partial g_{32}}{\partial x_2}, \\
\left( \nabla \cdot \mathbf{Q} \right)_{(3)} &= \frac{1}{g_2 g_{32}}\frac{\partial Q_{33}}{\partial x_3},
\end{split}
\end{equation}

\begin{equation}
\textstyle
\mathbf{Q}:\nabla \bm{v} =  l^2 \rho \nabla \cdot \bm{v} \frac{1}{3} \left[
  \left(\nabla \bm{v}_{(11)}-\nabla \bm{v}_{(22)}\right)^2
+\left(\nabla \bm{v}_{(11)}-\nabla \bm{v}_{(33)}\right)^2
+\left(\nabla \bm{v}_{(33)}-\nabla \bm{v}_{(22)}\right)^2 \right],
\end{equation}
where $g_2 = r$, $g_{31}=r$, and $g_{32}=\sin \theta$ are the metric components for the spherical coordinates, $(x_1, x_2, x_3) = (r, \theta, \phi)$ (see again \citet{stone}). 
These equations are discretized as
\begin{equation}
 \begin{split}
 \left( \nabla \cdot \mathbf{Q} \right)_{1,i,j,k} &= 
 \frac{g2b^2_i g31b_i Q11_{i,j,k} - g2b^2_{i-1} g31b_{i-1} Q11_{i-1,j,k}}{g2a^2_i g31a_i dx1b_i}, \\
 \left( \nabla \cdot \mathbf{Q} \right)_{2,i,j,k} &= 
 \frac{g32b^2_j Q22_{i,j,k} - g32b^2_{j-1} Q22_{i,j-1,k}}{g2b_i g32a_i^2 dx2b_j}
 + \frac{Q11_{i,j,k} + Q11_{i,j-1,k}}{2 g2b_i g32a_j}\left(\frac{\partial g32a_j}{\partial x_2}\right), \\
\left( \nabla \cdot \mathbf{Q} \right)_{3,i,j,k} &= \frac{Q33_{i,j,k}-Q33_{i,j,k-1}}{g2b_i, g32b_j dx3b_k},
\end{split}
\end{equation}

\begin{equation}
\begin{split}
\textstyle
(\mathbf{Q}:\nabla \bm{v})_{i,j,k} =  &\textstyle l^2_{i,j,k} d_{i,j,k} (\nabla \cdot \bm{v})_{i,j,k} \frac{1}{3} [
[(\nabla \bm{v}_{(11)})_{i,j,k}-(\nabla \bm{v}_{(22)})_{i,j,k}]^2 \\
&+[(\nabla \bm{v}_{(11)})_{i,j,k}-(\nabla \bm{v}_{(33)})_{i,j,k}]^2 
+[(\nabla \bm{v}_{(33)})_{i,j,k}-(\nabla \bm{v}_{(22)})_{i,j,k}]^2], 
\end{split}
\end{equation}
where $g2a$, $g31a$, and $g32a$ are $g_2$, $g_{31}$, and $g_{32}$ defined on the``a-mesh", whereas 
$g2b$, $g31b$, and $g32b$ are those defined on the``b-mesh". $dx1b_i$, $dx2b_j$, and $dx3b_k$ represent the width of the ``b-mesh" 
at the $i$-, $j$-, and $k$-th grid points in the $r$, $\theta$, and $\phi$ directions, respectively. Finally the velocity-gradient tensor 
$(\nabla \bm{v}_{(11)})_{i,j,k}, (\nabla \bm{v}_{(22)})_{i,j,k}$, and $(\nabla \bm{v}_{(33)})_{i,j,k}$ are given by the following equations:

\begin{equation}
\begin{split}
\left(\nabla \bm{v}_{(11)} \right)_{i,j,k} = &\frac{v1_{i+1,j,k} - v1_{i,j,k}}{dx1a_i},\\
\left(\nabla \bm{v}_{(22)} \right)_{i,j,k} = &\frac{v2_{i,j+1,k} - v2_{i,j,k}}{g2b_i dx2a_j}+
\frac{v1_{i,j,k}+v1_{i+1,j,k}}{2g2b_i}\left(\frac{\partial g2b_i}{\partial x_1}\right), \\
\left(\nabla \bm{v}_{(33)} \right)_{i,j,k} = &\frac{v3_{i,j,k+1}-v3_{i,j,k}}{g2b_i g32b_j dx3a_k}+
\frac{v2_{i,j,k}+v2_{i,j+1,k}}{2g31b_i g32b_j}\left(\frac{\partial g32b_j}{\partial x_2}\right)+
\frac{v1_{i,j,k}+v1_{i+1,j,k}}{2g31b_i}\left(\frac{\partial g31b_i}{\partial x_1}\right).
\end{split}
\end{equation}

\section{Numerical Convergence Tests \label{check}}
\begin{figure}[hbt]
\begin{picture}(0,200)
\put(10,10){\includegraphics[width=75mm]{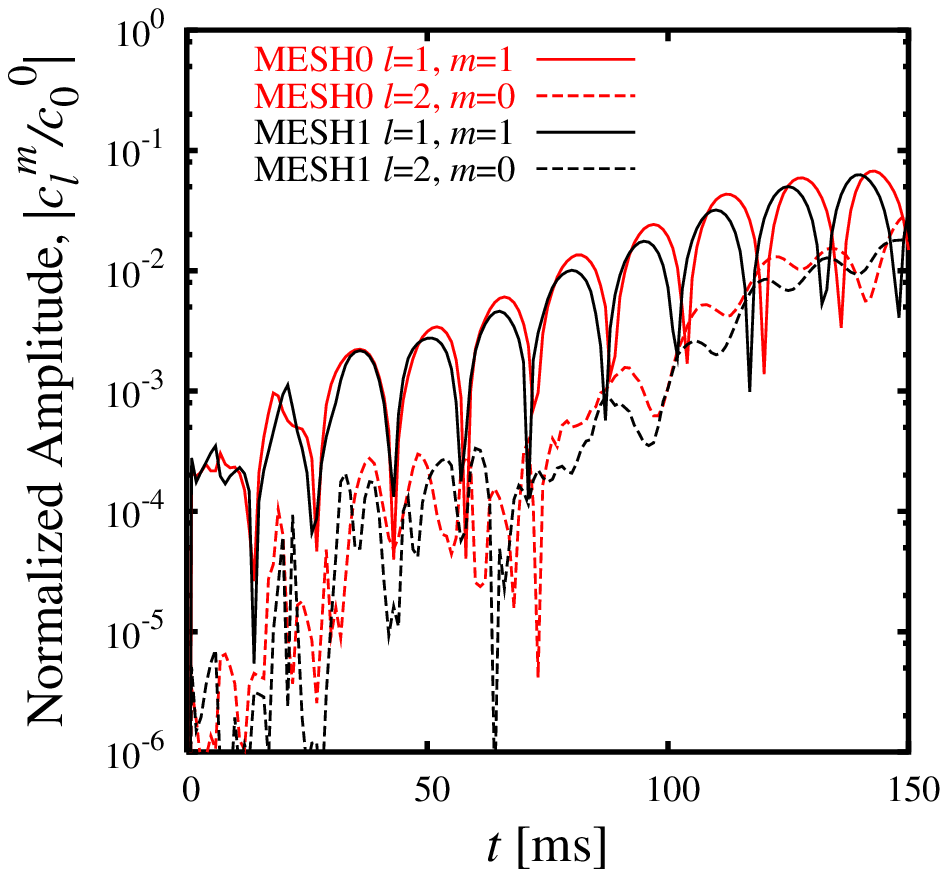}}
\put(10,0){\scalebox{1.1}{\footnotesize (a) single-mode $l=1, m=1$ perturbation}}
\put(235,10){\includegraphics[width=75mm]{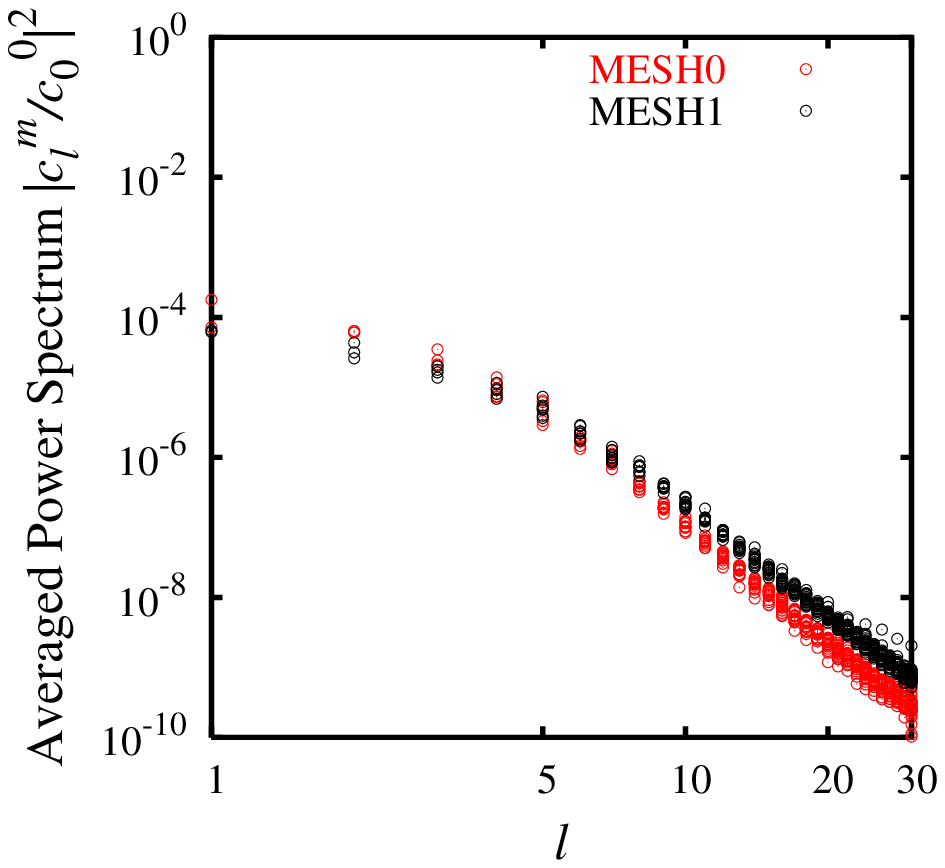}}
\put(240,0){\scalebox{1.1}{\footnotesize (b) multimode random perturbation}}
\end{picture}
\caption{The convergence tests. (a) The time evolutions of the normalized amplitudes $|c^m_l/c^0_0|$ in the linear phase. The single-mode perturbation with 
$l=1, |m|=1$ is imposed initially. The models with $300 \times 30 \times 60$ and $300 \times 60 \times 120$ mess points are referred to 
as MESH0 and MESH1, respectively. (b) The time-averaged power spectra $|c^m_l/c^0_0|^2$. The average is taken over 
$150 \leq t \leq 400$~ms. In this comparison, the random multi-mode perturbation is imposed.}
\label{fig16}
\end{figure}

In order to see if the numerical resolution employed in the main body is sufficient, we increase the number of angular grid points and compare the results. 
Figure~\ref{fig16} (a) shows the time evolutions of the normalized amplitudes $|c^m_l/c^0_0|$ in the linear phase. In this comparison, we impose the 
$l=1, |m|=1$ perturbation initially. We refer to the model with $300 \times 30 \times 60$ mess points as MESH0 and to that with 
$300 \times 60 \times 120$ grid points as MESH1in the figure. It is found that the linear growth rates agree with each other reasonably well although the coarser
mesh slightly overestimates the growth time. 
Figure~\ref{fig16} (b) presents the power spectra $|c^m_l/c^0_0|^2$ that are time-averaged over the nonlinar phase. The random perturbation is 
imposed in this case. It is again clear that the results for MESH0 is in good agreement with those for MESH1.

It should be mentioned that for MESH0 it takes 32 parallel processors about 1.5 days to compute the evolution up to $t=400$ ms, while MESH1 needs 22 days 
even for 128 parallel processors. This is partly because of the difference in the Courant numbers, which are set to be 0.5 for MESH0 but to be 0.1 for MESH1 
to better use the tensor-type artificial viscosity. Although this severe limit of CPU time does not allow us to do more thorough convergence tests, 
we think, based on the results of the tests shown above, that our results given in this paper are credible.

\clearpage

\input{tab1.tex}

\clearpage

\clearpage

\clearpage

\clearpage

\clearpage

\clearpage

\clearpage

\clearpage

\clearpage

\clearpage

\clearpage

\clearpage

\clearpage

\clearpage

\clearpage

\end{document}

%% file: tab1.tex
\begin{table}[h]
\begin{center}
\caption{Summary of all the models.}
\vspace{4mm}
\begin{tabular}{clc}
\hline
\hline
  Model & \hspace{18mm} Perturbation & Nuetrino Luminosity $L_\nu$\\
  & &  [$10^{52}$ ergs s$^{-1}$] \\
\hline
  I  & single-mode, $l=1, m=0$& 6.0\\
  I\kern-.1emI  & multi-mode, $l=1, m=0$ and $l=1, |m|=1$ & 6.0\\
  I\kern-.1em I\kern-.1em I & single-mode, $l=1, m=0$ with & 6.0 \\
& random perturbation at $t=400$ ms & \\
  I\kern-.1emV & random perturbation & 6.0 \\
  V& random perturbation & 6.4\\
  V\kern-.1emI & random perturbation & 6.8\\
  V\kern-.1emI\kern-.1emI & axisymmetric random perturbation & 6.0 \\
  V\kern-.1emI\kern-.1emI\kern-.1emI  & axisymmetric random perturbation & 6.8 \\
   I\kern-.1emX & none & 6.0 \\
  X & none & 6.8 \\
\hline
\hline
\end{tabular}
\label{model}
\end{center}
\end{table}